\documentclass[crcready]{iosart2x}

\newcommand{\Fref}[1]{Figure~\ref{#1}}

\newcommand{\Eref}[1]{Equation~\ref{#1}}

\newcommand{\Sref}[1]{Section~\ref{#1}}

\usepackage{xcolor}
\usepackage{paralist}
\usepackage{todonotes}

\pubyear{0000}
\volume{0}
\firstpage{1}
\lastpage{1}

\begin{document}
\makeatletter
\let\put@numberlines@box\relax
\makeatother

\begin{frontmatter}

\title{Towards end-to-end verifiable online voting: adding verifiability to established voting systems}
\runningtitle{Towards end-to-end verifiable online voting}

\author[A]{\fnms{Mohammed} \snm{Alsadi}\ead[label=e2]{m.alsadi@surrey.ac.uk}},
\author[B]{\fnms{Matthew} \snm{Casey}\ead[label=e3]{m.casey@pervasive-intelligence.co.uk}},
\author[A]{\fnms{Constantin Catalin} \snm{Dragan}\ead[label=e4]{c.dragan@surrey.ac.uk}},
\author[C]{\fnms{Fran\c{c}ois} \snm{Dupressoir}\ead[label=e5]{f.dupressoir@bristol.ac.uk}},
\author[D]{\fnms{Luke} \snm{Riley}\ead[label=e6]{l.riley@kcl.ac.uk}},
\author[E]{\fnms{Muntadher} \snm{Sallal}\ead[label=e7]{msallal@bournemouth.ac.uk}},
\author[A]{\fnms{Steve} \snm{Schneider}\ead[label=e1]{s.schneider@surrey.ac.uk}},
\author[A]{\fnms{Helen} \snm{Treharne}\ead[label=e8]{h.treharne@surrey.ac.uk}},
\author[F]{\fnms{Joe} \snm{Wadsworth}\ead[label=e9]{m.alsadi@surrey.ac.uk}} 
and
\author[G]{\fnms{Phil} \snm{Wright}\ead[label=e10]{philip.wright@civica.co.uk}}
\runningauthor{S. Schneider et. al.}
\address[A]{Surrey Centre for Cyber Security, \orgname{University of Surrey}, \cny{UK} \printead[presep={\\}]{e1,e2,e4,e8}}
\address[B]{Pervasive Intelligence, \cny{UK} \printead[presep={\\}]{e3}}
\address[C]{Department of Computer Science, \orgname{University of Bristol}, \cny{UK} \printead[presep={\\}]{e5}}
\address[D]{Department of Informatics, \orgname{King's College London}, \cny{UK} \printead[presep={\\}]{e6}}
\address[E]{\orgname{Bournemouth University}, \cny{UK} \printead[presep={\\}]{e7}}
\address[F]{formerly Electoral Reform Services Ltd., \cny{UK}} 
\address[G]{Civica Election Services, \cny{UK} \printead[presep={\\}]{e10}}


%
%

\begin{abstract}
Online voting for independent elections is generally supported by trusted election providers. Typically these providers do not offer any way in which a voter can verify their vote, and hence the providers are trusted with ballot privacy and in ensuring correctness. Despite the desire to offer online voting for political elections, this lack of transparency and verifiability is often seen as a significant barrier to the large-scale adoption of online elections. Adding verifiability to an online election increases transparency and integrity, as well as allowing voters to verify that the vote they cast has been recorded correctly and included in the tally. However, replacing existing online systems with those that provide verifiable voting requires new algorithms and code to be deployed, and this presents a significant business risk to commercial election providers. In this paper we present the first step in an incremental approach which minimises the business risk but demonstrates the advantages of verifiability, by developing an implementation of key elements of a Selene-based verifiability layer and adding it to an operational online voting system. Selene is a verifiable voting protocol that publishes votes in plaintext alongside a voter's tracker. These trackers enable voters to confirm that their votes have been captured correctly by the system, such that the election provider does not know which tracker has been allocated to which voter. This results in a system where even the election authority running the system cannot change the result in an undetectable way, and gives stronger guarantees on the integrity of the election than were previously present. We explore the challenges presented by adding a verifiability layer to an operational system.  We describe the results of two initial trials conducted within real contested elections, which obtained that survey respondents found this form of verifiability easy to use and that they broadly appreciated it.  We conclude by outlining the further steps in the road-map towards the deployment of a fully trustworthy online voting system.
\end{abstract}


\end{frontmatter}

\section{Introduction}\label{sec:intro}

Verifiability in electronic voting (e-voting) plays an important role in contributing to the trust in electronic voting systems through offering both voters and observers an opportunity to verify independently whether votes have been recorded and tallied correctly.  Numerous verifiability schemes have been proposed in the literature both for polling-place electronic voting, (e.g.~\cite{pretavoter,scantegrity,civitas,wombat}) and for remote voting~\cite{helios,belenios,pgd,selene}.  However, although some current commercial internet voting systems may contain verifiability mechanisms, they typically do not provide full end-to-end verifiability, as  they  do  not  provide proofs  or confirming  evidence that supports clear individual and universal verifiability.

Within the UK, Civica Election Services (CES)\footnote{previously Electoral Reform Services (ERS)} are the leading provider of independent election services, including e-voting—\(42\%\) of ballots run by CES are electronic, and a further \(28\%\) are mixed mode, combining electronic and postal votes. Their system is used for organisational ballots, such as for trades unions, political parties, professional societies and building societies, and is used by 4 million individual voters a year. Currently in the UK, e-voting is not allowed for statutory (for example, for strike action) or political ballots. Yet e-voting is advocated as a way of improving the engagement in the electorate, but only if it can be demonstrated to be sufficiently secure~\cite{political:constitutional:2015}, something that a provable layer of security with verifiability can add. The CES online voting system provides voters with credentials that they can use to login to submit their ballots, with votes then stored securely within a database. Access to the online system is via a web browser. Once the voters have cast their votes, the votes are used in the tally according to the pre-defined rules for the election. Once a voter has cast their vote, they have no further access to the system and cannot in any way verify their vote or the election. CES are therefore trusted to collect, store and tally the votes, and to maintain the privacy of voters. This trust can be audited by client organizations, and by Government entities in some cases.

Our aim is to move real-world online elections towards full end-to-end security and verifiability to reduce the trust that needs to be placed in providers and improve transparency. We aim to do so in steps, for two reasons.
First, we aim to slowly get general voters to expect electronic elections to provide certain features, such as individual and public verifiability. This reflects the observation by Kulyk et al.~\cite[p. 18]{DBLP:conf/interact/KulykHRV19} that ``the idea of verification, being a fairly alien concept, is problematic. More needs to be done to familiarize voters with the differences between paper and Internet voting.''
Second, although in principle providing a complete implementation of a new fully end-to-end secure and verifiable system from scratch would be the ideal approach, their use in real elections is not something that can be easily achieved, especially with certification constraints,\footnote{For example, CES are a Secretary of State-appointed ``scrutineer'' for industrial ballots, and changes to their systems could put that appointment at risk if not done carefully.} when a trusted but not end-to-end-verifiable system already exists.
We therefore take a more pragmatic, layered approach, which we describe below. In this paper, we focus on the first step on that roadmap: the addition of a verifiability layer to support the verification of the ballot's integrity independently of the election provider. As a consequence, the election provider themselves—or malicious insiders or external attackers gaining access to their critical systems—cannot change the result without risking detection.

\subsection{Roadmap to Fully End-to-End Secure Electronic Voting}\label{sec:intro:roadmap}

As mentioned, the work presented here represents only the first step towards transforming existing \emph{trusted} electronic voting systems into \emph{trustworthy} electronic voting systems.
This roadmap will help us justify some of our protocol and design choices. We highlight, in the rest of the paper, situations in which considering only the addition of verifiability would lead to different, more natural choices.

However, we do note that the roadmap itself provides an opportunity for a progressive exposure of the public to the value—and cost—of privacy and verifiability in electronic and online voting.
Indeed, each step is designed to provide additional benefit to the end user (the voter) while keeping each successive change to a minimum. Each of these steps can—in isolation—be the subject of usability experiments, or the topic of some public communication about the desirable properties of electronic voting systems, and of voting systems more generally.

\paragraph*{Step 1: Independent Verifiability} This paper proposes a verification layer, deployed independently of the voting system being used, and without requiring any changes to it. This verification layer is not trusted for privacy, and the voting system is not trusted for verifiability—although they must not collude.
Deploying this layer may help make visible to end users the potential of verifiable voting, and also supports conducting usability studies on real, large contests.

\paragraph*{Step 2: Casting Encrypted (and optionally signed) Ballots} In a next step, the voting front-end would allow the election provider to collect encrypted ballots directly—signing could be added at the cost of some interaction with the verification layer during ballot casting.
Although this is a very small change on paper, when compared to the previous step, it is a significant change for a business: the election provider's ability to produce the election result now depends on continued engagement from the verification layer. Further, it also requires that the casting of electronic ballots be done through a website with the ability to run code on the voter's machine—a constraint which may disenfranchise some of the existing online voters.

These constraints, combined with the practical need to also support mixed mode elections—where some ballots may be cast by post, lead us to consider this a discrete step, requiring a different form of stakeholder engagement. Practical deployment of this step on real elections—of the kind we consider here—would also require much more careful software and systems engineering.

\paragraph*{Step 3: Voter Credentials} In a third step, after having introduced users to the value of privacy and verifiability in electronic—and perhaps postal—voting, we would introduce them to their own credentials. In this step, we would—--in a way similar to Belenios--—generate and store voter credentials as part of the system, and give them sufficient information to recover them upon casting their ballot and verifying the election.
Deployment of this step would make verification more transparent—allowing users who wish to truly independently verify the public ballot and their own vote without relying on the verification layer's infrastructure, but would not increase the usage burden overmuch.

Deploying this step securely, and without losing the privacy and verifiability value of the first two steps, will require careful engineering to make credential management independent from both the voting provider and the verification layer—for example through the use of a third party.

\paragraph*{Step 4: Voter Registration} The next step is naturally to allow those users who wish to generate their own credentials and register them for use in a ballot to do so on their own.
Here, the main challenges to solve will involve the risk of disenfranchisement due to the loss of credentials, and the risk to privacy and verifiability related to the compromise of end devices \emph{outside} of the time window where the election takes place.

We expect here a very slow build-up of the proportion of voters who would engage with the generation and registration of their own credentials. However, this could be accelerated by the growth of other uses of cryptography in public life: the requirements here will be similar to those for an electronic wallet or electronic identity card.

Once we have reached this step, in which voters are generating and managing their own keys, then we have reached the stage where the system provides end-to-end verifiability.

\paragraph*{Step 5: Coercion Mitigation} Finally, with voters aware and in control of their cryptographic credentials, coercion-mitigation measures (i.e., coercion-evidence, coercion-resilience or coercion-resistance) can be deployed.
Deploying such measures expands the type of elections the voting system can be used in. In some settings (where coercion takes specific forms, or in settings where alternative protocol choices are possible), it may be possible to deploy them earlier in the roadmap.

As discussed, we now focus mainly on the first step—the addition of verifiability—and recall the constraints associated with the next steps when they are relevant to discussions.

\subsection{Existing End-to-End Secure Electronic Voting}
Adapting existing e-voting systems to incorporate verifiability requires significant change to the structure and voting experience of these systems, and in particular may affect usability.  Many existing 
schemes that support verifiability require some voter action such as cut-and-choose at casting time \cite{benaloh, pretavoter,helios,belenios,wombat} or code voting  \cite{DBLP:conf/voteid/HelbachS07,DBLP:journals/compsec/JoaquimFR13,pgd}, making the act of casting the ballot more complex for voters and therefore less desirable for commercial election providers. 
For example Acemyan et al. \cite{DBLP:conf/uss/AcemyanKBW14} show that only \(43\%\) of voters were able to verify their vote successfully with the Benaloh challenge across Prêt-a-Voter, Helios, and Scantegrity II, which obtained a low usability score.
Kulyk et al. \cite{DBLP:conf/interact/KulykHRV19} show that code-based verification is easier to verify than Benaloh-challenges when comparing the two, but identify some usability issues with both approaches.

\emph{Tracker-based} approaches to verification instead allow voters to check that their plaintext vote is included in the final tally—attached to a tracker whose value they know, without breaking the ballot's privacy or requiring much additional work on the voter's part during voting. Examples include a Boardroom approach~\cite{DBLP:conf/voteid/ArnaudCW13}, the sElect scheme~\cite{DBLP:conf/csfw/KustersMST16}, and the Selene scheme~\cite{selene}.
Arnaud, Cortier and Warinschi~\cite{DBLP:conf/voteid/ArnaudCW13} propose a very simple tracker-based system, however this is not receipt-free.
sElect provides an elegant and lightweight mechanism which requires a tracker (which they call a ``verification code''') to be provided at the point of voting for subsequent verification. It is appropriate for low-coercion elections, providing vote privacy and verifiability provided the voter does not reveal their tracker during the election.
Selene has additional mechanisms for mitigating the risk of coercion.  The data needed for verification is provided to voters after the election and not at the time of voting. Each voter a unique tracker which is only revealed to them in a coercion-mitigating way after the voting period has closed. The allocation of trackers to voters is not known to the election provider.

A more recent approach to verification is based on \emph{return codes}~\cite{DBLP:conf/voteid/KhazaeiW17}—as used in SwissPost's proposed system~\cite{swisspost}, which are used to obtain confirmation from the voter that their ballot is being recorded as cast.
This approach is promising, as it does not require any onerous verification-related work to be done by the voter while casting their vote, and could likely be integrated with the CES ballot-casting workflow. We do note, however, that the approach does not provide plaintext verification, requires an independent device, and cannot be deployed post hoc. This makes it less directly suitable for deployment on mixed mode elections if verifiability is desired also on the postal results.

\subsection{Layering verifiability on a commercial system}
To introduce verifiability to a commercial e-voting system, while reducing the risk that any real election can go wrong as a result of the implementation, the selected scheme must be capable of being layered on top of the system to ensure recoverability of the election.
Layering reduces the impact on the existing system because it can remain mostly unchanged meaning that less regression testing and auditing is required in deploying the new verifiability mechanism—including for settings in which auditing is imposed by regulations, rather than a simple internal requirement.
As is the nature of experimental software, in the advent of a failure in the verifiability layer, the election can fallback to the existing system to complete the election with little impact on the voters so that the election result can still be produced.
Such layering could still have an impact on privacy if implemented badly—we discuss this in \Sref{sec:analysis}.

Summing up our constraints, we want to layer verifiability on top of CES's system in a way that:
\begin{enumerate}
    \item does not require changes to the current election system (to increase the chance that paying clients will opt into participating in the research);\label{itm:changes}
    \item does not jeopardize the correct running of the elections it is deployed on (to protect CES's organizational reputation);
    \item minimizes risk to privacy (in particular, provides privacy against the verifiability layer);
    \item does not require the voter to take any verification-related steps at the time they cast their ballot;
    \item fits with our roadmap, and in particular can later be extended to support full end-to-end verifiability and ballot privacy (including coercion-mitigation, and privacy against a malicious insider); and eventually support mixed-mode voting.\label{itm:roadmap}
\end{enumerate}

Without constraint~\ref{itm:roadmap}, it would in fact make sense to choose the simplest appropriate approach to deploy verifiability with private ballots, and deploy a sElect-like mechanism for ballot verification.
However, the requirement that our chosen verifiability layer can later be extended also with strong ballot privacy—including coercion-mitigation—leaves us with a choice between Selene and return codes. Rønne et al.~\cite{electryo} show how Selene can be deployed on paper ballots, while Iovino et al.~\cite{DBLP:conf/fc/IovinoRRR17} add a verifiability layer to JCJ using Selene. Selene therefore seems easier to extend to mixed mode voting than return codes, which involve a casting-time interaction.

In this paper, we therefore describe how Selene verifiability can be added as an external layer to an existing, leading, commercial internet voting system, without significant change to the voters' experience during ballot casting, or to the overall structure of the system. Our design is engineered to better support further incremental extensions towards full end-to-end verifiable and coercion-mitigating electronic or mixed mode elections.
The VMV demonstrator described in this paper is not a full implementation of Selene, but rather a stepping stone towards it to enable initial trials with voters and evaluate the implementation's impact, particularly on the voter experience. While it does provide the full cryptographic functionality of the Selene verification mechanism, it also executes some of the cryptographic operations in Selene that should be the responsibility of the voters themselves. We make clear where VMV differs from pure Selene below. 

\subsection{Contributions}\label{sec:intro:cont}

Our main contributions in this work are as follows:
\begin{enumerate}
	\item We develop an approach which can be used to add verifiability to some established online voting systems.
	\item We provide an implementation of the Selene protocol, highlighting particular points where practical implementations must refine its specification.
	\item We evaluate the impact on voters of providing verifiable e-voting in real-world elections.
	\item We define a roadmap, sufficient to modify an established e-voting system and give full end-to-end verifiability and privacy.
\end{enumerate}

\subsection{Paper Organisation}\label{sec:intro:po}

The paper is organised as follows. The rest of this introduction covers core concepts of electronic voting, as well as related work beyond that already covered. In~\Sref{sec:selene} we describe the Selene protocol, highlighting the implementation challenges that it presents. The VMV demonstrator is then described in~\Sref{sec:architecture}, including how our Selene implementation deviates from the original protocol in order to overcome its challenges. We provide an overall analysis of the demonstrator in~\Sref{sec:analysis}.  In~\Sref{sec:trials} we describe the operation of the VMV demonstrator and the results from real-world trials, including feedback from voters. Finally in~\Sref{sec:conclusion} we conclude by reflecting on our findings and defining how the limitations of the demonstrator may be overcome to provide a full end-to-end secure and verifiable election system which can be implemented with minimal impact on existing systems.

\subsection{Electronic Voting}
A voting system provides a way of enabling voters to make a secret selection from among a number of choices, to submit their selection or vote into the system, and for the system to calculate a choice (winner or winners) from all the votes cast based upon the rules of the election.  The election is administered by the Election Authority who is responsible for managing the list of eligible voters (the electoral roll) and ensuring that only they can vote. The Election Authority is also responsible for tallying the votes and for calculating the result of the election.  The voting system should protect the secrecy of the ballot so that it is not possible to tell how any particular voter voted, even if the voter wishes to show that they voted in a particular way (e.g. if they are selling their vote or being coerced).  It should also provide the election authorities with the correct result.  These requirements must be assured even in the face of adversaries who may wish to manipulate or subvert the system to deliver a particular result even if that does not reflect the votes cast.  Electronic systems provide particular challenges because the capture, transmission and processing of electronic votes can be subject to cyber attack: an attacker might attempt to alter or delete votes, add additional votes, or interfere with the tally. Traditional paper-based elections have processes to counter these kinds of attacks, relying on physical properties of paper and ballot boxes, and electronic voting systems must also have mechanisms in place to defend the integrity of the election.

\subsubsection{Verifiability}

End-to-end verifiability for electronic voting systems \cite{DBLP:journals/corr/BenalohRRSTV15} provides a way to provide assurance in the integrity of the election result. Mechanisms are provided for enabling all of the steps in the process from vote casting through to vote tallying to be {\em verified} so that incorrect processing of the votes at any stage can be detected, meaning that an attacker cannot undetectably alter the result of an election.  Such verifiability is made up of {\em individual verifiability}, where voters check that the system does correctly contain their vote, or its absence if they did not cast a vote (only the voter can check this since only the voter knows their vote), {\em universal verifiability}, where certain steps, such as the tallying of the result, can be checked by any independent party, and other considerations such as the absence of {\em clash attacks}. We do not provide formal definitions here as we are inheriting the verifiability properties from Selene. {\em Eligibility verifiability} provides a check that only votes from eligible voters have been included. Verifiability focuses on information produced by the election system, so it is the election data produced by the system that is verified rather than the election system itself. It is important that this data preserves ballot anonymity, and does not give away how any voter voted or provide a voter with the means to prove how they voted.

Individual verifiability consists of two components. The first is {\em cast as intended}, which verifies that the vote submitted into the system really does reflect the voter's intention.  While this is immediately evident for a paper vote, it is not always obvious for devices that capture the vote for submission, particularly if the vote has been encrypted or encoded in some way.  The second is {\em recorded as cast}, which verifies that the vote recorded or held by the electronic ballot box does indeed match the vote that was submitted by the voter.  Universal verifiability is concerned with the correctness of the results calculated from the recorded votes---this is typically known as {\em counted as recorded}.  If the voting system provides verification mechanisms for all three of these steps, then it is known as {\em end-to-end verifiable}.

\subsection{Related Work}\label{sec:intro:related}

There are numerous proposals for end-to-end verifiable voting systems, including Pr\^et \`a Voter \cite{pretavoter}, Wombat \cite{wombat}, Scantegrity II \cite{scantegrity}, Helios \cite{helios}, Belenios \cite{belenios}, Civitas \cite{civitas}, and Selene \cite{selene}. These make use of common verifiability mechanisms to underpin the integrity of the election.  Some use paper whereas others are purely electronic, and some are intended for use in the polling place, whereas others are intended for remote voting from a voter's own device. 

A common mechanism for verifying that a vote is {\em cast as intended} is the Benaloh Challenge \cite{benaloh}, a cut-and-choose method for confirming that a vote has been constructed correctly.  After the vote has been created for submission to the election system, typically by encrypting it, the voter can choose whether to cast the vote or to audit it. An audit involves revealing the vote and providing the evidence that it was indeed encrypted correctly.  Audited votes cannot also be cast since their contents have been revealed, contrary to the secret ballot.  Therefore a voter can audit several votes before finally deciding to submit an unopened one.   Audits are similar to random sampling of votes: if votes are not constructed correctly then an audit would catch this, so an election with sufficient successful audits gives some level of evidence that all cast votes are also constructed correctly.  This approach is taken in Helios, Belenios, Civitas and Wombat.  This cut-and-choose approach is also applicable to pre-constructed ballot forms, which can either be used to vote or can be audited (without a vote) to check they have been constructed correctly.  This approach is taken in Pr\^et \`a Voter and Scantegrity II.

Typically a voter will retain a record of the vote that was cast, and will be able to confirm that this matches the published list of all the votes cast, to verify {\em recorded as cast}.  Most commonly this record will contain the vote in some encrypted form, so it does not reveal the vote. A system that does not provide the voter with a way of revealing how they voted is known as {\em receipt-free}. Receipt-freeness is a desirable requirement of voting systems.

An alternative approach to obtaining individual verifiability is through the use of Code Voting as provided for example in Pretty Good Democracy \cite{pgd}.  This approach provides (by post or some other private channel) each voter with a code sheet which contains a voting code for each candidate, and a return code.  The voter casts a vote by submitting the code for their candidate, and on receiving the return code they obtain confirmation that the vote has been correctly received, since only the election system has knowledge of the voting codes and return codes.  This verifiability property assumes that the codes remain secret, at least until the verification step has taken place.

The approach taken by Selene is different again: at the end of the election all votes are published alongside a tracker, and each voter is provided privately with their tracker.  They are then able to confirm directly that the vote against their tracker is indeed the vote they cast, verifying that it has been cast as intended.  This approach enables voters to verify their votes in the clear, rather than in encrypted or code form, so much of the design of Selene is to protect privacy, by ensuring that voters cannot prove their tracker to any other party, and that they do not obtain it until after all the votes have been published.

\emph{Universal verifiability} works with the published list of encrypted votes, which can be processed in a universally verifiable way to obtain the result of the election.  There are two main approaches to achieving this.  The first is to use an anonymising mix-net \cite{chaum:mixnet,sakokilian:mixnet} to shuffle and re-encrypt the ballots, resulting in a list of encrypted ballots that cannot be matched to the voters' receipts.  Zero-knowledge proofs of shuffling or randomised partial checking enable independent verification that this has been carried out correctly.  The resulting list is decrypted to reveal the plaintext votes, which can then be tallied publicly in the normal way.  All of these steps obtaining the result can be independently verified. This is the approach taken by Pr\^et \`a Voter, Wombat, Helios v3 and Civitas, as well as Selene.

An alternative approach is to encrypt the votes in such a way that we can make use of homomorphic encryption, enabling the encrypted total for each candidate to be obtained from the individual encrypted votes.  These encrypted totals can then be decrypted to reveal the results, without revealing any individual votes.  All of these cryptographic steps can also be carried out in a verifiable way.  This is the approach taken in Helios and Belenios.

\paragraph*{Large-Scale Deployments of Cryptographic Online Voting} We are not the first to aim at large scale deployments of online voting schemes in real elections. Other deployments on potentially similar scales are Government-driven.

Norway allowed the use of online voting~\cite{DBLP:conf/voteid/Gjosteen11} in some elections, but stopped in 2014. The short deployment was used to support further research on the effectiveness of verifiability at detecting tampering~\cite{DBLP:journals/adt/GjosteenL16}.

Estonia also uses online voting to complement their citizens' ability to vote in polling stations. Their system provides natural coercion resistance by allowing re-voting (so that a coercer would need to actively keep coercing the voter throughout the election period) and giving precedence to votes cast in-person. Their deployment started from a core private but not end-to-end verifiable system and more recently added verifiability features~\cite{DBLP:conf/ev/HeibergW14,DBLP:conf/voteid/HeibergMVW16}. It is backed by existing infrastructure for the management of voter credentials (through Government-issued electronic identification), which our system cannot assume exists.

A number of Swiss cantons have been experimenting with electronic voting for parts of their monthly \emph{votations}. A variety of systems have been deployed in practice, some based on Benaloh challenges, and some on code voting. A more recent proposal by Scytl and SwissPost~\cite{swisspost} suggests the use of return codes to avoid Benaloh challenges while providing cast as intended verification. This same system is also in use in Australia and in some French elections.
No documentation is available on interactions of this voting mode and pen-and-paper-based voting for the same elections.
\section{Selene Overview}\label{sec:selene}

Selene~\cite{selene} is an end-to-end verifiable election cryptographic protocol which is designed to allow voters to verify their votes without significant impact on the voting process, and to make individual verifiability easy to understand (avoiding, in particular, the need for Benaloh challenges). Here the underlying verification cryptography can remain hidden to help improve the voter experience. Selene achieves this by allocating each voter with a unique tracker such that the election provider does not know which voter has which tracker, and that each voter only knows their tracker after the election tally, when they can verify their vote. The former is to ensure voter privacy, while the latter is designed to provide coercion mitigation.   

The protocol consists of five stages: 
\begin{inparaenum}
	\item election set-up, including the generation of the election keys and encrypted trackers,
	\item generation of tracker commitments and distribution to voters of the first part of their commitment,
	\item voting, where encrypted and signed votes are recorded for a voter with their encrypted tracker,
	\item mixing and decryption, where all of the encrypted tracker and vote pairs are securely shuffled then decrypted, and
	\item notification of trackers where each voter receives the information required for them to recover their tracker and hence verify their vote.
\end{inparaenum}

Throughout these stages, Selene relies upon the use of a mix-net~\cite{Chaum:1981:UEM:358549.358563,Ramchen:2010:PSA:1924892.1924904}, a distributed system that securely and verifiably shuffles its input data according to some unknown permutation. Shuffling within the mix-net takes advantage of an encryption scheme which can re-encrypt data without first decrypting it (such as ElGamal~\cite{elgamal}). In our implementation we use Verificatum \cite{verificatum} to provide our mix-net and we use their $(k,t)$-threshold ElGamal Encryption System  to generate the election public encryption key and to share the decryption key.   

Initially, each node is used to create an encryption key pair, with each node holding a share of the private key, enforcing the need for a threshold number of nodes to be involved in decryption. The network of nodes can then be used to randomly shuffle tuples encrypted using the corresponding public key. During the shuffle, each node in turn receives a list of encrypted tuples, randomises this list and then re-encrypts each tuple before passing the new list onto the next node in the network. Since each node is designed to be operated by an independent organisation, no single node can be used to match the original order of the input tuples with the final order of the output tuples.   The output tuples can thus safely be decrypted using a threshold of the nodes. In this way, a list of tuples of encrypted trackers and votes can be shuffled and decrypted such that there is no correspondence between voters and the output plaintext trackers and votes. Verificatum~\cite{verificatum} is a popular mix-net implementation proven in public elections, and which provides provably secure cryptography to achieve the above operations, and hence is used in VMV.

Selene also relies upon an append-only web bulletin board (WBB), to broadcast data in such a way as to guarantee that data—once broadcast—cannot be amended or deleted. (The mix-net also makes use of a separate WBB in order to communicate data privately between nodes.) In Selene, the WBB is used to provide a record of the election data for verification and audit. This includes publishing, for example, the election public key, encrypted trackers, encrypted votes and shuffled plaintext votes. Since distributed ledgers provide a way of storing data in such a way that it can be append-only and recorded across multiple nodes run by independent organisations (much like the mix-net), they are a natural choice to provide a WBB, as implemented, for example, in vVote~\cite{CSF:CulSch14}.

\subsection{Election Set-up}\label{sec:selene:setup}

During the set-up stage of the election, the threshold encryption key pair is generated by the mix-net such that each node holds a share of the private key corresponding to the public key: \((sk_{T_1},\ldots,sk_{T_t}, pk_T)\), where \(t\) is the number of nodes in the mix-net, known as `tellers' to follow election terminology, and \(T\) refers to the whole mix-net so that \(sk_{T_i}\) is the private key share held by teller \(i\) and \( pk_T\) is the public key for the whole mix-net. The keys are generated using suitable parameters, including key length and such that \(sk_{T_i}, pk_T \in G\) where \(G\) is a cyclic group of order \(p\) with generator \(g\), and \(p\) is a suitably large prime number.

Next a sufficient number of unique random trackers \(n_i\) are generated for the \(n\) voters.  VMV generates these trackers, with access only to the number of voters (i.e., the number of trackers it is required to generate).  For VMV we used 8-digit numbers selected randomly without repeats.  The list of trackers is published so the lack of repetition is verifiable.
Each tracker is then encrypted using the election public key. ElGamal encryption requires that any data being encrypted must be within the cyclic group \(G\), and hence each tracker is first mapped to an element of \(G\) by calculating \(g^{n_i}\) before it is encrypted, to give \(n\) tuples
\begin{equation}\label{sec:selene:setup:eq}
(n_i, g^{n_i}, \{g^{n_i}\}_{pk_T})
\end{equation}
where \(i \in \{1,\ldots,n\} \), \(g\) is a generator of the group \(G\) and \(\{m\}_{pk_T}\) represents encryption of \(m \in G\) using the public key \(pk_T\). As proposed in the description of Selene we use trivial randomness for this encryption to enable verification. 

Once all of the tracker tuples have been generated, each of the \(\{g^{n_i}\}_{pk_T}\) encrypted terms are passed to the mix-net to be randomly shuffled (and re-encrypted) to give the list of encrypted trackers \(\{g^{n_{\pi(i)}}\}'_{pk_T}\) in an unknown order. Here, \(\pi(i)\) refers to the random ordering of the \(i\)th element, while \(\{\ldots\}'\) represents re-encryption. Now that the trackers are in an unknown random order, they can be allocated arbitrarily to each voter since there is no way that a \(\{g^{n_{\pi(i)}}\}'_{pk_T}\) can be mapped back to its corresponding \(\{g^{n_i}\}_{pk_T}\) (under the discrete log computational hardness assumption and the assumption that at least one of the mix nodes is honest).

During this stage, to enable independent verification of the election parameters, each of the \(p\), \(G\), \(g\), \(pk_T\), \((n_i, g^{n_i}, \{g^{n_i}\}_{pk_T})\) and \(\{g^{n_{\pi(i)}}\}'_{pk_T}\) terms are published to the WBB, along with the proofs of correct shuffling produced by the mix-net.

\subsection{Generation of Tracker Number Commitments}\label{sec:selene:commitments}

To ensure that each voter can recover their unique tracker at the end of the election, while also ensuring that the tracker allocation is not changed once allocated, each voter is given a commitment value \(\beta_i\) linked to their tracker. At the end of the election, the pair of commitment values \((\alpha_i, \beta_i)\) form an ElGamal ciphertext encrypted under the voter's public key \(pk_i\), the plaintext of which is their tracker. Here the generation of \(\beta_i\) relies upon the multiplicative homomorphic properties of ElGamal.

Each teller $j$ generates \(n\) random numbers \(1 \leq r_{i,j} < p\) where \(i=1,\ldots,n\), for \(j=1,\ldots,t\). These are used to form the values:
\begin{equation}\label{sec:selene:commitments:eq}
\alpha_i = g^{r_i} \mbox{ and } \beta_i = {pk_i}^{r_i}.g^{n_{\pi(i)}}, \mbox{ where } r_i = \sum_{j=1}^{t} r_{i,j}.
\end{equation}

The voter receives only $\beta_i$ before the voting phase, and will be allowed to compute $\alpha_i$ after the notification state (\Sref{sec:selene:notification}) as the values $g^{r_{i,j}}$ for \(i=1,..,n\) are held privately by each teller $j$.

The values $\beta_i$ and $\alpha_i$ are not passed in clear, as that would defeat the purpose of keeping the trackers hidden from the voters and malicious parties until the end of the voting phase. Thus, each teller $j$ produces an encryption under the election public key for each voter $i$: 
$(\{{g^{r_{i,j}}} \}_{pk_T}, \{{pk_i}^{r_{i,j}} \}_{pk_T})$ 
together with a non-interactive zero-knowledge proofs of knowledge (NIZKPoKs) that it is well-formed, i.e., the same randomness $r_{i,j}$ is used \cite{selene}. 
Each teller provides a decryption under its own key of $\{{pk_i}^{r_{i,j}}.g^{n_{\pi(i)}} \}_{pk_T}$ together with a NIZKPoKs of correct decryption \cite{selene}. Then, the decryptions are combined to compute  ${pk_i}^{r_{i,j}}.g^{n_{\pi(i)}}$.  The ciphertext $\{{pk_i}^{r_{i,j}}.g^{n_{\pi(i)}} \}_{pk_T}$ is easily obtained by homomorphically multiplying $\{{pk_i}^{r_{i,j}} \}_{pk_T}$ and $\{g^{n_{\pi(i)}} \}_{pk_T}$. These encrypted and decrypted values together with their NIZKPoKs are published on the WBB.


\subsection{Voting}\label{sec:selene:voting}

In Selene the voting stage is designed to minimise the impact on the voter's experience so that they are not required to perform any additional, potentially onerous, steps in order to vote, or to perform any verification at the time of voting. Therefore, when a voter casts their ballot, they form their vote \(\text{Vote}_i\), encrypt it using the election public key \(\{\text{Vote}_i\}_{pk_T}\) and sign it using their own signature key \(\text{Sign}_{V_i}(\{\text{Vote}_i\}_{pk_T})\).  (Selene also allows a version without signatures on votes.)  In addition, a NIZKPoK \(\Pi_i\) is generated to prove that the plaintext vote was encrypted correctly. In order to form \(\{\text{Vote}_i\}_{pk_T}\), the \(\text{Vote}_i\) must be mapped to a number in the cyclic group \(G\), as required by ElGamal encryption.

It is intended that these steps: voting, encryption (including NIZKPoK) and signing can all be achieved on the voter's own device. Since each step requires only relatively simple modulo arithmetic using large numbers, they can be achieved, for example, using JavaScript in a web browser or using a mobile app using bespoke software, all of which can be scrutinized by third-party auditors. Once these terms have been generated, they are sent to the election provider and recorded on the WBB as the tuple:
\begin{equation}\label{sec:selene:voting:eq}
(pk_i, \{g^{n_{\pi(i)}}\}'_{pk_T}, \beta_i, \text{Sign}_{V_i}(\{\text{Vote}_i\}_{pk_T}), \Pi_i)
\end{equation}

Note here that the election provider does not therefore know the content of the voter's vote since it is encrypted. Furthermore, the vote is recorded with its proof of knowledge guaranteeing that the content of the encryption is correct, and signed by the voter to allow for later verification.

Each of these stages so far therefore requires each voter to have a signature key pair used to sign their vote, together with an encryption key pair which is used to recover their tracker (see~\Sref{sec:selene:commitments} and~\Sref{sec:selene:notification}).

\subsection{Mixing and Decryption}\label{sec:selene:mixing}

Once the voting period has closed, the encrypted tracker and vote pairs from each of the voter's tuple (\Eref{sec:selene:voting:eq}) are extracted and then shuffled by the mix-net to form the tuples
\begin{equation}\label{sec:selene:mixing:eq1}
(\{g^{n_{\pi(i)}}\}'_{pk_T}, \{\text{Vote}_i\}'_{pk_T})
\end{equation}
These can now be decrypted by the mix-net, and each \(g^{n_{\pi(i)}}\) can then be mapped to its corresponding \(n_{\pi(i)}\) to form the plaintext tracker and vote pair
\begin{equation}\label{sec:selene:mixing:eq2}
(n_{\pi(i)}, \text{Vote}_i)
\end{equation}
which can subsequently be verified by voters. These tuples, together with the corresponding proofs of decryption are published to the WBB.

\subsection{Notification of Tracker Numbers}\label{sec:selene:notification}

The last stage of the Selene protocol is to allow voters access to their \(\alpha_i\) commitment value so that they can recover their tracker. Recall that \((\alpha_i, \beta_i)\) are the ElGamal ciphertext for the tracker encrypted under the voter's public key \(pk_i\). An ElGamal ciphertext is of the form
\begin{equation}\label{sec:selene:notification:eq1}
(g^r, pk^r.m)
\end{equation}
where \(g\) is the generator for the group, \(m\) is the message (\(m \in G\)), \(pk\) is the public key, \(r\) is a random value \(1 \leq r < p\), and "." is multiplication within the group. Here then, from~\Eref{sec:selene:commitments:eq} we want to form
\begin{equation}\label{sec:selene:notification:eq2}
(\alpha_i, \beta_i) = (g^{r_i}, {pk_i}^{r_i}.g^{n_{\pi(i)}})
\end{equation}
so that \(m = g^{n_{\pi(i)}}\) is the tracker mapped to a number in the group \(G\). To achieve this we therefore need to form \(g^{r_i}\), and because of the multiplicative homomorphic properties of ElGamal, \(r_i = \sum_{j=1}^{t} r_{i,j}\). Each teller $j$ therefore sends each voter $i$ the value \(g^{r_{i,j}}\). The voter's device can then multiply all of these values to form \(g^{r_i}\) and gain \(\alpha_i\), then decrypt the ciphertext using their private key \(sk_i\) to extract \(g^{n_{\pi(i)}}\), which can then be mapped to their tracker \(n_{\pi(i)}\) using the data published on the WBB.

Once the voter has their tracker, they can verify their plaintext vote on the WBB by using their tracker to find their vote. Note that Selene also allows alternative \(\alpha_i\) values to be generated by a voter (using their private key) to give a different tracker to provide some coercion mitigation, but we do not explore this mechanism in this paper—leaving it to Step 5 of our roadmap (see Section~\ref{sec:intro:roadmap}).

\subsection{Refining Selene for Implementation}\label{sec:selene:challenges}

While Selene offers two key advantages over other schemes which enable it to be used as a layer over an established online voting system, namely the separation of verification data generation to occur before and after the election, and the minimal impact on the voter experience which only modifies casting of ballots to require the additional encryption and signing, it does present other challenges which must be overcome in order to implement it for use in an operational system.  Selene has not previously been implemented, and several of these observations only became apparent to us as a result of our implementation.

These are:

\subsubsection{Voters are assumed to have their own trapdoor and signing key pairs suitable for use with Selene (\Sref{sec:selene:commitments} and~\Sref{sec:selene:notification})}\label{sec:selene:challenges:keys}
With a voter holding their keys and using the election public key, they submit only encrypted and signed votes to the system, thus guaranteeing vote privacy; and they use their trapdoor key to verify their vote after the election. However, this means that suitable key pairs must be generated in some way by each voter (or on their behalf) with their private keys held securely and only accessible by them. This in turn assumes either some sort of trusted authority who can provide the keys, or that the voters have sufficient knowledge and/or equipment to generate and store the keys themselves. The keys must also be kept safe for the full election period; if their private trapdoor key is lost, they cannot verify their vote. In general, this assumption cannot easily be met as it either requires a sufficient level of digital literacy in voters, or some sort of authority (such as a government) to provide the necessary infrastructure for voters.  With respect to the signing keys, in fact having the votes signed is optional in Selene: it provides a mechanism for checking eligibility of the votes cast and making it easier to detect ballot stuffing (i.e. by using the credentials of voters who have not voted to include votes for them) without relying on non-voters verifying that no ballot has been cast in their name.  Even without signatures such ballot stuffing could be detected by the voters concerned.  

\subsubsection{During voting (\Sref{sec:selene:voting}), it is assumed that the election provider can work only with encrypted votes}\label{sec:selene:challenges:encrypted}
While this is an important underpinning privacy-preserving requirement for the protocol, when applying Selene as a layer over an existing system which deals with plaintext votes, this requirement may need to be reconciled with recoverability of the election in the event of a failure of the verification layer, so the ability to provide an election result is not placed in jeopardy. Here, with only encrypted votes, recovery would require decrypting all of the submitted votes, so this will need to be made possible even if the verification layer has failed. Note that this is an issue for full end-to-end security in all established e-voting systems.
In an election where obtaining the election results takes priority over privacy protections, as is enforced by regulations in some of the elections CES is certified to scrutinize, the ability to recover from such failure would be required. Collecting encrypted votes may then need to be complemented with controlled recovery mechanisms (such as a key escrow, or an offline hardware security module) to ensure a result is obtained.
Systems such as Helios and Belenios, which prioritize privacy, offer no such recovery mechanisms—in those systems, the loss of the election credentials or of enough trustees to prevent threshold decryption would instead require running the election again.
	
\subsubsection{During voting, it is left open when submitted vote tuples (\Eref{sec:selene:voting:eq}) should be recorded on the WBB}\label{sec:selene:challenges:realtime} This can be either when they are received by the election provider, or after voting has finished.   
In a system where voters have the option to cancel their existing vote and vote again, the real-time recording to the WBB of submitted votes will require cancellation of submitted votes also to be on the record.  Selene does not consider this possibility explicitly, and assumes that votes posted on the WBB are votes to be tallied.  Posting votes during the election also provides real-time information concerning turnout, which may not be appropriate to publish, depending on the requirements of the election.  Here we avoid these challenges by only publishing submitted votes after the end of the polling phase.

\subsubsection{ElGamal cannot be used to encrypt arbitrary text (votes) without first mapping the text to a member of the cyclic group \(G\) (\Sref{sec:selene:voting})}\label{sec:selene:challenges:elgamal}
In order to map a vote to a member of the group, the full set of possible votes is therefore required in advance of voting. For yes/no votes or votes for a single condidate this is clearly straightforward, but for preferential voting, where the voter may be required to numerically order any number of candidates, this can result in millions of possible vote options.  Pre-defining and distributing these options then becomes difficult because of storage and communication requirements. It will instead be necessary to manage preference lists as tuples of candidates, adding complexity to the processing of the votes. 

\subsubsection{During the notification of trackers (\Sref{sec:selene:notification}), each teller individually sends their recorded value $g^{r_{i,j}}$ value to each voter}\label{sec:selene:challenges:notification}
This requires each teller to know enough information about the voter to send them securely and individually their randomness value via a secure channel. For example, does each teller therefore know every voter's email address (assuming email is sufficiently secure), or does each voter have access to every teller to retrieve their values? Since each teller is meant to operate independently, this may require personal information to be distributed to multiple third-parties and needs to be in some way constrained by appropriate privacy laws, such as the General Data Protection Regulation (GDPR)~\cite{gdpr}.  Alternatively the randomness values can be collated and forwarded on via a separate entity, for example as is proposed in Electryo \cite{electryo}.  In our case the strict separation between the CES system (which manages the connections with the voters) and the VMV system (where the tellers are situated) ensures that the VMV components will never learn voters' emails or any other personal information.

Lastly, while the protocol is well defined in~\cite{selene}, some elements, such as the choice of mix-net and WBB, are not defined, typically because these are necessary implementation detail. Nonetheless, suitable choices are required to ensure that the protocol is not compromised, while also ensuring that the public user interfaces of, for example, the WBB promote easy access to the published information to allow voters to verify their votes.

\section{Architecture and Implementation} \label{sec:architecture}

As described, our approach is to enhance an existing system with a verifiability layer in order to understand how such systems can be improved. Having chosen Selene as a suitable protocol to provide verifiability, in this section we describe both the existing system and how Selene is layered over it.

\subsection{Legacy System}\label{sec:architecture:legacy}

\begin{figure}[th]
	\centering
	\includegraphics[width=0.45\textwidth]{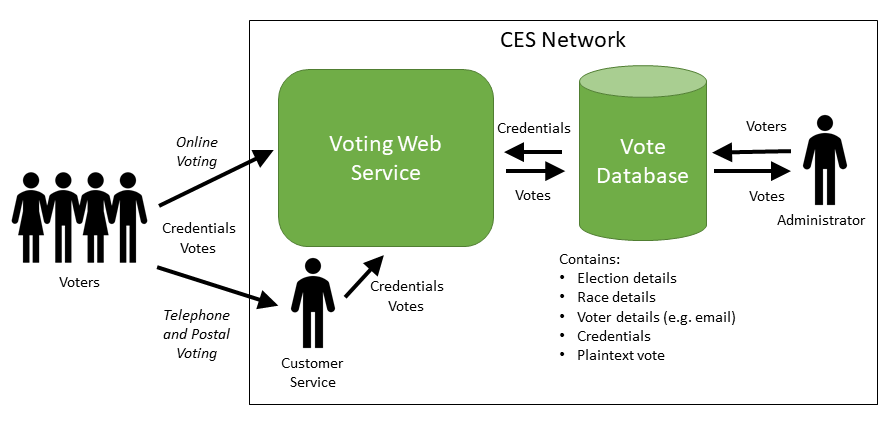}
	\caption{CES System Architecture} \label{sec:architecture:ers}
\end{figure}

Our chosen commercial partners are CES, who run a significant number of elections within the UK using a web-based voting system. Their system allows for online, postal and telephone voting to take place for up to hundreds of thousands of voters, with all cast ballots recorded in a relational database (\Fref{sec:architecture:ers}). This database is accessed via a secure web service operating from their data centres on Microsoft Windows servers.  When an election is to be run, CES receive details of the election type, voters, materials needed to be released to voters, such as candidate information, and tallying rules. Each election may consist of multiple races (different questions upon which each voter may vote). These data are then used to provision a suitable section of the CES web service which can be accessed by voters using their credentials when voting is opened. The database contains a list of all of the voters, their credentials and record of their vote (empty prior to voting).

When voting is opened, voters receive their security credentials which enable them to log into the web service via any supported browser using HTTPS and cast their ballot. These credentials may be received via email or by post. Once a voter has cast their ballot, the corresponding database table is updated to contain the plaintext (machine readable) vote for every race in which the voter has voted. Postal and telephone votes may be received for certain elections, in which case the corresponding votes are recorded within the same database table by customer service representatives. Furthermore, it is possible for a voter to cancel their vote, say by telephone, in order to re-vote if they wish.

At the end of the election, the tallying rules are applied to the collected votes and the results released to the commissioning organisation. CES do not make the election results public as it is the responsibility of the organisation to release the results depending upon their own rules.

This system therefore places complete trust within CES to run the election. This includes notifying all voters, providing the election materials and credentials, maintaining the integrity of the collected votes, tallying the votes using the correct rules, and releasing the corresponding results to the commissioning organisation. Since votes are held in plaintext and there is no way for a voter to verify their vote, trust is placed in the security of the system and the integrity of the staff. The use of plaintext votes is integral to the system in order to perform the tally.

\subsection{Design}\label{sec:architecture:design}

In implementing a verifiability layer with Selene, there are two overriding requirements: 
\begin{inparaenum}
	\item to provide individual and universal verifiability of the election, and
	\item to ensure that the established system remains intact in case the verifiability layer fails.
\end{inparaenum}
This latter requirement is driven by business need: when operating a large-scale, commercial election which is trialling experimental software, there must be a mechanism whereby the election can be easily recovered without loss of data.  Indeed this requirement dictates that the storage of plaintext votes and existing tallying mechanism remain as-is while the software is at the experimental stage and undergoing trials. Yet despite this, by adding voter verification and publishing the election results publicly, the election becomes transparent and, importantly, verification is able to expose any malicious change in the election result, thus reducing the required trust in the election provider. Once the experimental software is proven and made sufficiently robust for production use, then the requirement to maintain the existing system is removed.

As a consequence, in order to impact the least on the existing system, the design enforces the separation of the CES and VMV software, which is achieved simply by interfacing VMV via the relational database, which holds all of the plaintext votes, and by providing a separate user interface for vote verification and election auditing. 
By keeping the votes in plaintext within the CES system, and then adding verifiability to the voter record and their plaintext vote, the impact on the system is minimised because neither the voter user interface or processing need to change, while the desired verifiability can be added. Nonetheless, this compromise means that the existing lack of end-to-end privacy of votes within the CES system continues at this stage.

\begin{figure}[th]
	\centering
	\includegraphics[width=0.45\textwidth]{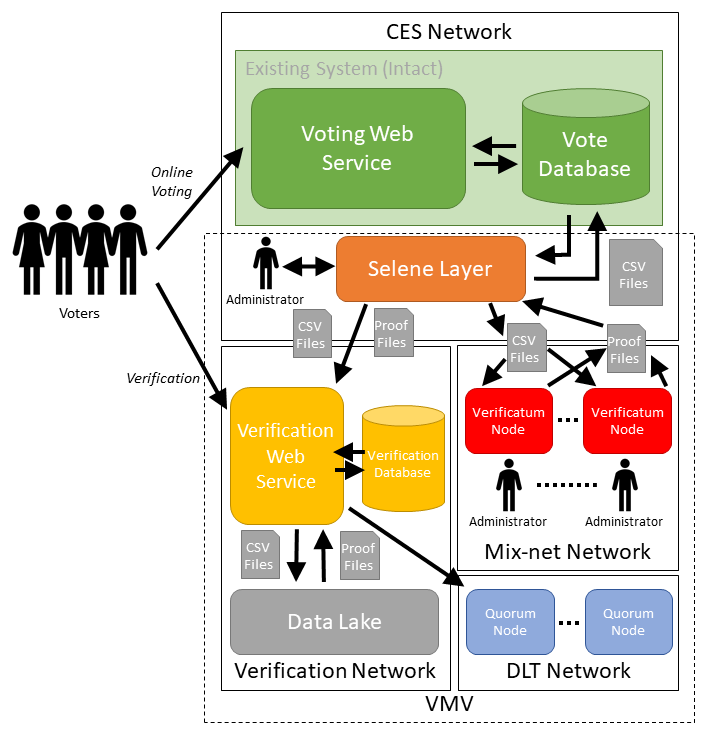}
	\caption{VMV System Architecture} \label{sec:architecture:vmv}
\end{figure}

The separated VMV software architecture is shown in~\Fref{sec:architecture:vmv}, which shows the relationship between the additional components and the CES system. The components in the architecture are:
\begin{description}
	\item[Voting Web Service] The existing CES e-voting system which operates without change except to provide additional information to voters to allow them to verify their vote.
	\item[Vote Database] The existing CES relational database holding all details about an election, voters and their plaintext vote (once a ballot has been cast). This is modified to add in the verifiability data per voter and is used as the input and output interface for VMV through the import and export of comma-separated values (CSV) data files.
	\item[CES Network] The secure network within which the Voting Web Service and Vote Database are held. Public access is only granted to the Voting Web Service within this network via HTTPS (and to vote only with credentials). Since the Selene Layer accesses voter and vote data, it is also run within the CES Network to ensure that all private data is kept securely within the network.
	\item[Selene Layer] Executes the Selene protocol by taking data from the Vote Database as CSV files, communicating with the Verificatum Nodes to perform shuffling and decryption, and with the Verification Web Service to publish verification data, including produced CSV and NIZKPoK proof files. These operations are initiated by an administrator using a computer running within the CES Network.
	\item[Verificatum] A series of independently-operated nodes running the Verificatum software. Two or more independent organisations can run a Verificatum Node which is initialised by the Selene Layer. Each Verificatum Node can communicate with each other node within the Mix-net Network. Prior to a mix-net operation, such as shuffling, each node is supplied with identical CSV input and produces identical CSV output together with the corresponding proof files~\cite{verificatum}.
	\item[Mix-net Network] Each Verificatum Node is run within its own secure network hosted by each independent organisation. Access to each Verificatum Node is only granted to the other Verificatum Nodes and the Selene Layer, which controls the Verificatum operations.
	\item[Verification Web Service] A web service with a user interface which allows administrators to publish verification data, auditors to view the published election data and voters to verify their vote. This forms the public face of the VMV demonstrator and allows published files to be served to users. Publication requires privileged access granted to administrators via user accounts. Only administrators have accounts, while anyone can view published data. 
	\item[Verification Database] Holds the data necessary to run the Verification Web Service, including administrator user accounts and an index of each election's verification data. This includes the list of the CSV and proof files held in the Data Lake, and their corresponding contract addresses in the Quorum cluster, such that they can be retrieved via the Verification Web Service.
	\item[Data Lake] Holds the published CSV and NIZKPoK proof files in a repository which is only accessed via the Verification Web Service.
	\item[Verification Network] A secure network in which the Verification Web Service and Data Lake operate. Public access is only granted to the Verification Web Service within this network via HTTPS.
	\item[Quorum Node] A series of independently-operated nodes running the Quorum software, a particular Distributed Ledger Technology \cite{quorum}. Two or more independent organisations can each run one or more Quorum Nodes. Each Quorum Node can communicate with each other node within the DLT Network. When a file is published via the Verification Web Service, it is saved to the Data Lake and a hash of the file is committed to the Quorum cluster. Periodically, the hash is verified against the file held in the Data Lake to ensure its integrity. 
\end{description}

All data held within the Vote Database, Verification Database, Data Lake, Verificatum Nodes and Quorum Nodes should be held resiliently such that they are backed up to prevent data loss. For example, each Verificatum Node holds privately within its file system its share of the election private key \(sk_{T_i}\).

Verificatum was chosen as the preferred mix-net implementation because it is open source, has worked successfully in a number of large-scale elections, has a proven cryptographic protocol, works well with ElGamal encryption and produces the desired NIZKPoK for each operation.

The Verificatum Nodes within the mix-net require a threshold number of the nodes to perform cryptographic operations. The number of nodes within the mix-net and the number required for a threshold is configurable when the mix-net is created. For example, four Verificatum Nodes can be run such that a threshold of three of them is needed to operate. This allows for one (and only one) node to be removed from the mix-net for it to still be able to operate. If less than the threshold number of Verificatum Nodes is available, then no mix-net operations can take place. Nodes may be removed from the mix-net through failure or if the operator of the node is thought to be compromised or malicious. Similarly for the Quorum cluster. 

Quorum was chosen as it is open source (based upon Ethereum~\cite{ethereum}) and being developed as an enterprise-ready DLT solution by J.P. Morgan~\cite{quorum}, ensuring that it is sufficiently robust for commercial operation.

\begin{figure}[th]
	\centering
	\includegraphics[width=0.45\textwidth]{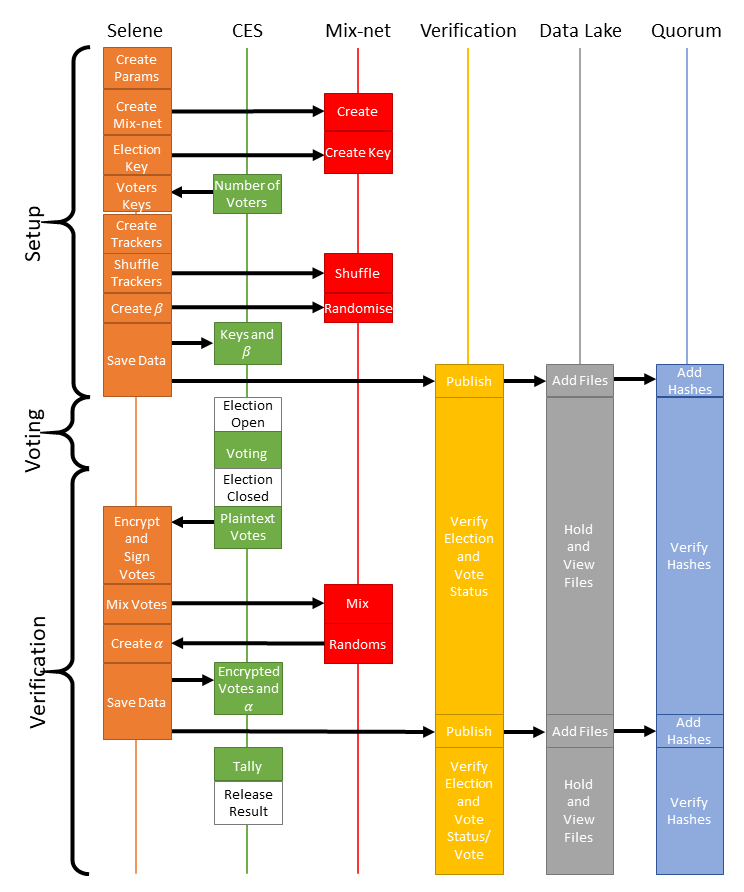}
	\caption{Interaction between Components During an Election} \label{sec:architecture:sequence}
\end{figure}

The sequence of interactions between the components is shown in~\Fref{sec:architecture:sequence}, and the following provides detail on each stage of operation, describing the various design choices associated with each component, while relating to the Selene protocol in~\Sref{sec:selene}.

\subsection{Setup}\label{sec:architecture:setup}

Once the election has been defined by CES, and the number of voters quantified, then the Selene Layer can be used to set up the election verifiability parameters.

The first stage is to initialise the cryptographic parameters ({\bf Selene: Create Params}) which will be used to create the corresponding encryption and signing keys (\Sref{sec:selene:setup}). For example, this includes the cyclic group \(G\) with order \(p\), and the associated generator \(g\). Once this has been done, the election key pair can be created. Recall that in Selene, the election encryption key pair is created by the mix-net so that shares of the private key \(sk_{T_i}\) are distributed across the mix-net nodes \(i\), while the public key \(pk_T\) is available for third-parties to encrypt data. In Verificatum, each node in the mix-net is initialised with data relevant to the election, including the cryptographic parameters, election name, and the IP address of each Verificatum Node in the mix-net ({\bf Selene: Create Mix-net}). For convenience, a copy of the Selene Layer software is used on each Verificatum Node to run the Verificatum commands necessary to initialise the local files ({\bf Mix-net: Create}). When this has been completed, the Selene Layer is used ({\bf Selene: Election Key}) to create the election key pair ({\bf Mix-net: Create Key}). During key pair creation, each Verificatum node communicates with each other node to form a consensus on the key and ensure that each has a share of the private key (with each share remaining private to the node).

We now turn to our first challenge in the Selene protocol (\Sref{sec:selene:challenges:keys}) where Selene assumes that all voters have their own encryption and signing keys initialised from suitable cryptographic parameters. In VMV voters do not have their own keys, even though the keys are assumed to be available to complete the Selene protocol. In an ideal scenario, each voter would have their own keys either allocated to them via organisational or national infrastructure, or they would be able to generate and store them securely themselves. This is not the case for the CES system, and hence during the setup of the election, the encryption \((sk_i, pk_i)\) and signing keys for all of the voters are generated for them ({\bf Selene: Voters Keys}) by taking from the CES system the number of voters in the election ({\bf CES: Number of Voters}). This is an unavoidable compromise in the protocol which means that the Selene Layer holds all of the keys for the voters and is therefore a trusted party. However, while the Selene Layer holds both the private and public keys for the voters, the private keys are not shared with the CES system so that it is not possible for CES staff to encrypt and sign votes --- only the Selene Layer can do this.

Once all of the keys have been generated, the next stage is to create the trackers ({\bf Selene: Create Trackers}). Random trackers are created (enough for every voter), which are then mapped to a number in the cyclic group \(G\) and encrypted using the election public key \(pk_T\) (\Sref{sec:selene:setup}). These encrypted values are then shuffled ({\bf Selene: Shuffle Trackers}) using the mix-net ({\bf Mix-net: Shuffle}) so that there is no correspondence between the input and output encrypted trackers from the mix-net.

Alongside the trackers, the first part of their commitments, \(\beta\), ({\bf Selene: Create \(\boldsymbol{\beta}\)}) are then generated (\Sref{sec:selene:commitments}) by first requesting that every Verificatum Node generates a random value $r$ for each voter ({\bf Mix-net: Randomise}). This random number generation is not part of Verificatum and is instead completed by the controlling Selene Layer on each node. These random values are then combined in encrypted form and transformed to be one half of the ciphertext of the tracker encrypted under each voter's public key \(pk_i\), as described in Equation~\ref{sec:selene:commitments:eq}.  The value $g^{r}$ is also held privately by the node.  This process also involves the decryption of data by the mix-net but for simplicity, we omit the details here as the process is fully described in~\cite{selene}.

The last stage of the election setup is to allocate each voter their encryption and signature key pairs, an encrypted tracker and the corresponding \(\beta\). CES is then provided with this data ({\bf Selene: Save Data}) to store in the Vote Database ({\bf CES: Keys and \(\boldsymbol{\beta}\)}) against each voter (without the private keys). All of the public data is then sent to the Verification Web Service for publication ({\bf Verification: Publish}). During publication, the Verification Web Service saves each of the supplied CSV and proof files to the Data Lake ({\bf Data Lake: Add Files}) then calculates a SHA-256 hash of each file and then commits this hash to the Quorum cluster ({\bf Quorum: Add Hashes}). Once committed, the files are made available for public viewing by the Verification Web Service ({\bf Verification: Verify Election and Vote Status}), which allows files to be retrieved ({\bf Data Lake: Hold and View Files}) and periodically checks the corresponding hashes against the contract in Quorum ({\bf Quorum: Verify Hashes}).

Our choice of using a Data Lake is motivated by the size of the files that are generated. With, for example, keys with 3027 bits, a voter record (\Eref{sec:selene:voting:eq}) consists of approximately 4000 bytes, so that with 1000 voters, the corresponding data file is just under 4 MB and 100,000 voters 400 MB. While Quorum is designed to immutably store data, the larger the amount of data that needs storing, the longer the required consensus protocols take to run across the Quorum Nodes. Consequently, a Data Lake is used to hold the files, which can then be of arbitrary size, while the Quorum cluster only holds a hash of each file. If any file is changed, its hash will therefore not match to that which is stored in Quorum. While this is not ideal since it requires the hash to be checked when retrieving any file to ensure its integrity, it is more practical without compromising on the immutability of the files and allowing independent verification of the hashes.

Data is committed to the Quorum cluster using the notion of a `contract'. A contract is similar in concept to an object in an object-oriented programming language. Each contract template is written using the Solidity~\cite{solidity} language, encapsulating data and methods which operate on the data. To commit data to the Quorum cluster, the compiled contract is loaded and a new instance created with the required data and/or methods executed. The contract is then committed to all nodes in the cluster through a consensus protocol so that it is written to the blockchain. Once committed, the address of the contract is returned, and this can be used to retrieve the contact and its data.

Once all of the data for this stage has been published, anyone can view the public data for the election via the Verification Web Service ({\bf Verification: Verify Election and Vote Status}). As part of the email sent out by CES to all voters with their security credentials, each voter is also sent their \(\beta\) value. This can be used by the voter to verify that their \(\beta\) exists within the verification data (their `Vote Status'). This therefore enables any interested third-party to verify, for example, the number of voters, that every voter has a unique \(\beta\), and to also independently verify all of the NIZKPoK.

\subsection{Voting}\label{sec:architecture:voting}

Once the setup is complete, CES may open the election for voting. Here, the CES system remains unchanged in that voters use their security credentials to login to the Voting Web Service (via HTTPS) and submit their vote ({\bf CES: Voting}). Each vote is recorded in plaintext within the Vote Database. CES also allow voters to cancel their votes via telephone, and then to re-vote.

Here we face two challenges presented by Selene which assumes the end-to-end encryption of votes (\Sref{sec:selene:voting}). First, votes are not end-to-end encrypted since voters submit their vote in plaintext which is then recorded in the database (\Sref{sec:selene:challenges:encrypted}). As discussed, this means that CES are trusted to maintain the privacy of votes, but this was required to allow the CES system to remain intact and recoverable in the event of a failure in VMV. However it would be a straightforward adaptation of the system to receive and manage only encrypted votes.

Second, votes are not recorded in real-time to the Quorum cluster (\Sref{sec:selene:challenges:realtime}). There are three reasons why this is not done:
\begin{inparaenum}
	\item voters may cancel their votes, and hence their vote record may change after it has been committed to the cluster,
	\item real-time recording needs a more tightly-coupled software integration in that VMV would have to perform various functions every time a voter voted, and this more tightly-coupled integration in the demonstrator was undesirable given that the established system would then have to be significantly modified, and
	\item when data is published in real-time to the DLT then information about real-time turnout is revealed which might not in general be permitted.
\end{inparaenum}

\subsection{Verification}\label{sec:architecture:verification}

Once the election period has ended and voting closed, the final set of verification data may be generated and the tally performed. First, all of the plaintext votes are exported from the Vote Database ({\bf CES: Plaintext Votes}). This is only done within the CES Network so that the plaintext votes are never compromised. Alongside the plaintext votes, the public keys and encrypted trackers stored within the Vote Database for each voter are also exported within the CSV file. This enables the Selene Layer to find the corresponding private signature key for each voter.

In order to encrypt all of the plaintext votes, we must now overcome the limitation of ElGamal which can only be used to encrypt numbers within the cyclic group \(G\) (\Sref{sec:selene:challenges:elgamal}). This can either be overcome by first supplying a complete list of all possible plaintext votes, or by finding all distinct votes which have been cast. The former is possible where production of the list of distinct plaintext votes is tractable, such as for yes/no votes or similar, but which becomes too time-consuming in preferential voting with lots of candidates. The latter can then be used to find all distinct votes from those that have been cast, which at the worst case will be as many as there are ballots cast. Once a list of distinct votes has been generated, each can be mapped to a unique number in the cyclic group for encryption. For example, for every vote option, a unique random number \(v_i\) is generated which can be mapped into a number within the cyclic group \(G\) with generator \(g\) to yield
\begin{equation}
V_i = g^{v_i} \;{\mbox{mod}}\; p
\end{equation}
where \(p\) is the prime order of the group.

With all of the votes mapped, the Selene Layer is then used to encrypt and sign the votes for each voter ({\bf Selene: Encrypt and Sign Votes}) to produce the completed vote record (\Eref{sec:selene:voting:eq}). This includes the proof of correct encryption. Votes are signed using the Digital Signature Algorithm (DSA)~\cite{fips:186:4}.

The encrypted tracker and encrypted vote tuples for each voter are then extracted for mixing (\Sref{sec:selene:mixing}). Each Verificatum Node receives the list of tuples ({\bf Selene: Mix Votes}) and is instructed to mix them ({\bf Mix-net: Mix}). This performs a shuffle of the tuples before decryption. The result is a list of tuples holding the plaintext tracker in the cyclic group and the corresponding plaintext vote. The trackers in the cyclic group are then mapped back to the trackers (\Eref{sec:selene:mixing:eq2}).

The final stage of the Selene protocol is to enable voters to calculate their tracker by supplying them each with their random values \(g^{r_{i,j}}\) held by the Verificatum Nodes. Unfortunately, providing each independent Verificatum Node with a means to form a secure channel to every voter is not practical, as this would require some form of identifying information to be given to each node (such as an email address or credentials for a voter to access the node securely). Here then we address this challenge (\Sref{sec:selene:challenges:notification}) by focusing instead on the end result, namely providing each voter with their \(\alpha_i\) commitment.

In Selene, the concept is that each voter must generate their \(\alpha_i\) from the random values provided to them. In some way this allows the voter to avoid coercion supposedly because they cannot be forced to reveal their true \(\alpha_i\), and can instead generate an alternative \(\alpha_i'\) which points to a different vote record. To achieve this, the values must be sent securely, and the voter must have the necessary software needed to calculate their \(\alpha_i\) or an alternative. Both of these impose constraints on the voter which detract from their experience and make verification more difficult.

Our approach is to assume that the voter can receive their \(\alpha_i\) directly without having access to their random values. This does increase the risk of coercion, but prevents the distribution of personal data to the Verificatum Nodes. Here then instead, each Verificatum Node shares the random values with the Selene Layer running within the CES Network ({\bf Mix-net: Randoms}), which then calculates the \(\alpha\) values ({\bf Selene: Create~\(\boldsymbol{\alpha}\)}). The \(\alpha\) values can then be distributed by CES to the voters without revealing any personal data to VMV.

The encrypted vote, corresponding signature and \(\alpha\) are provided to CES ({\bf Selene: Save Data}) to store in the Vote Database ({\bf CES: Encrypted Votes and~\(\boldsymbol{\alpha}\)}), and so that the \(\alpha\) can be provided to each voter once the tally has been completed ({\bf CES: Tally}). This is achieved by sending them an email with their \(\alpha\) and \(\beta\) embedded within it, together with instructions on how to verify their vote.

Here also, all of the public data is then sent to the Verification Web Service for publication ({\bf Verification: Publish}). Once committed, the files are made available for public viewing by the Verification Web Service ({\bf Verification: Verify Election and Vote Status/Vote}).
\begin{description}
\item[Software Environment]\label{sec:architecture:software}  The implementation of the VMV demonstrator consists of five software environments:
\item[CES] The CES software runs intact on Microsoft Windows servers in their data centres. The interface with VMV is through the import and export of CSV files from the Vote Database.
\item[Selene and Verificatum] For portability, the Selene Layer was built in Java, using the Bouncy Castle cryptographic API~\cite{bouncycastle}. Verificatum is also built using Java~\cite{java}. By using Java, both pieces of software can be run on Windows or Linux servers providing flexibility for deployment, while Java also provides strong support for cryptography.
\item[Quorum] Quorum is written in Go~\cite{go} and can be deployed to Linux servers.
\item[Verification]
The Verification Web Service was built using Ruby-on-Rails~\cite{rubyonrails} to support rapid application development of a web service which requires both a user interface and an underlying infrastructure that supports the Data Lake and access to Quorum. A web service can therefore be accessed by any internet-enabled computer and a suitable web browser, and promotes the use of best-practice guidelines for the development of user interfaces to promote a good voter experience. The web service can be run on Linux servers with the Data Lake provided by suitable resilient storage (for example Amazon Web Services S3~\cite{aws}).
\end{description}

The source code for the VMV software \cite{VMV} is open source under the MIT licence, and is available at https://github.com/saschneider/VMV.  Version 1.0, used in the system described in this paper, is archived at 10.5281/zenodo.3695909.

\section{Security Analysis}\label{sec:analysis}

VMV crucially adds verifiability to the CES system. This means that the trustworthiness of CES can be audited.  However, while it maintains the CES system's privacy it does not currently provide coercion-mitigation on its own; instead CES provide a manual process to cancel a vote which can be used for coercion-mitigation.  However, this is less of a concern because the CES ballots conducted were all considered to be low coercion elections.  We now discuss more generally the security properties of VMV with the architecture presented in this paper, comparing them to those of Selene.

\subsection{Verifiability and Verification}
The Selene protocol is designed to provide end-to-end verifiability: voters can check that their vote was \emph{cast as intended} and \emph{recorded as cast} (individual verifiability), and any independent person can check—from publicly available information—all of the cryptographic operations performed by the mixers and tellers (universal verifiability).  For example, verifying the requirements that the trackers should be unique through to validating the proofs of shuffling and decryption have all been carried out correctly.

VMV inherits from Selene its universal verifiability property, since all of the required information is published on the ledger, which serves as an append-only web bulletin board.

However, individual verifiability can only be obtained in a weaker model. Indeed, VMV manages voters' private keys on their behalf, and does not give voters as much control over the opening of trackers as a pure implementation of Selene would. This design choice was made to simplify the user experience, removing the need for voters to create, register and manage their own private keys ahead of the election. 

Still, Selene's verifiability properties carry over to VMV under the assumption that VMV is trusted to\footnote{Our specific architecture somewhat mitigates the consequences of violations of this trust assumption; we discuss this in Section~\ref{ssec:distributed}.}:
\begin{enumerate}
\item protect and use voters' private keys (trapdoor and signing keys) only as
  specified by the protocol; and
\item provide the correct $\alpha$ and $\beta$ terms to each voter based upon
  their identifier.
\end{enumerate}
We note in particular that VMV is \emph{not} trusted to maintain the integrity of the verifiability data it holds (in the data lake); instead, the integrity of that data is assured by the hash values held in the ledger, under a weaker distributed trust assumption. In our particular scenario, neither CES nor VMV control a threshold of the ledger nodes.

Under this simple trust assumption, and since VMV provides voters with the ability to decrypt tracker commitments after the election, VMV equips the CES system with verifiability properties it does not have on its own. In particular, CES no longer needs to be trusted for the integrity of the election, and cannot modify or stuff ballots without risking detection.

\subsection{Ballot Privacy}
In the existing CES system vote privacy is handled by human and business processes, and data protection measures within CES.  However, CES do hold sufficient information to tell how the voter voted.  VMV does not change this since the votes are still collected in the same way prior to being encrypted for the remaining verification process.  To obtain complete ballot privacy, end-to-end encryption of the ballots would required. Given the experimental nature of this VMV development, a design choice was made not to encrypt votes end-to-end so that ballots could be recovered in the event of a catastrophic failure.  However the ultimate aim once the system is no longer experimental would be to have the votes encrypted on the client side before submission, so that CES only ever handle encrypted votes.

However, VMV is \emph{not} trusted for ballot privacy. Indeed, VMV has no access to the voter identities beyond the pseudonyms (code numbers allocated by CES), and crucially does not have access to the election authority's private key, under which all sensitive information published to the ledger is encrypted. Selene already provides the assurance that the published verifiability information does not compromise the privacy of the vote, under a standard independence assumption on the mixing nodes. This is maintained by VMV.

In fact, even if VMV and a threshold of the tellers are compromised (and the adversary can thus decrypt ballots and mixed trackers) then only the relationship between trackers and pseudonyms can be exposed, without exposing the further link between pseudonyms and voters. Indeed, only CES know the relation between voters and their pseudonyms (this is essential for communicating with the voters). Thus, VMV is not trusted for ballot privacy.


The system is designed to maintain the existing levels of vote privacy—which places trust in the election provider. However, we do note that our system as presented does not implement coercion-mitigation: the election provider passes on enough information to voters for VMV to point them to their vote on the WBB. This information is therefore sensitive for privacy.
In practice, VMV has to be involved in the recovery, and the service could be turned off in the event of a catastrophic leak. Further, a closer integration with the election provider would also better support private vote verification by requiring that the voter confirms to VMV, and interactively with the election provider that they are indeed the voter whose tracker commitment is being queried, without revealing their identity or pseudonym to VMV. We could not implement this additional authentication step—which requires interaction with a commercially sensitive database—in an uncertified research prototype.

\subsection{Receipt-Freeness}
Receipt-freeness means that the system should not provide evidence to the voter that they can subsequently use to prove how they voted. When Selene's verification layer is deployed on a receipt-free voting scheme, the resulting voting scheme is also receipt-free, since the additional information contained in the tracker, its commitment, and its encryption to the voter as an $(\alpha,\beta)$ pair does not allow a voter to prove how they voted. Indeed, possession of the trapdoor allows a voter to open the $\beta$ term publicly associated to them to any tracker of their choice, as long as the $\alpha$ term is sent to them through an untappable and deniable channel.

While this is the ultimate goal, the current implementation does not provide receipt-freeness because
\begin{inparaenum}[(1)]
\item we do not provide a way for voters to generate a fake $\alpha$, and
\item it is impractical to provide an untappable channel.
\end{inparaenum}
All the communication between CES and the voters is via email, which we are assuming is not tapped in practice (but could be used to prove the $\alpha$ value). We chose not to allow voters to obtain fake trackers, as we felt making this feature available would require voter education alongside it ahead of the election.  This would interfere with our experimental goal to assess VMV's initial usability, and the voters' understanding of verifiability and its impact on transparency and trust.  Introducing the generation of Selene fake trackers will be appropriate at a future stage of VMV where the communication between VMV and the voter is more secure.

\subsection{Coercion: Threats and Mitigation}
Selene provides some level of coercion-mitigation in that $\alpha$ values can be faked by voters as protection against coercion, however it does not prevent coercers from directly observing a voter casting their ballot or receiving the $\alpha$ value from the tellers. Some discussion of how these issues can be incorporated into Selene has been provided in \cite{selene}, including the use of Civitas-style fake credentials \cite{civitas}, or providing the tellers with the information for fake $\alpha$s to send back.  VMV does not implement any of these mechanisms, and therefore does not provide coercion-mitigation.  A coercer can ask a voter for their $\alpha$ value in order to ensure that the coerced voter has voted as instructed.  The current version of VMV  therefore assumes a low-coercion election.

In practice the system run by CES also allows for votes to be cancelled through a manual process, removing them from the database of votes, and allowing another vote. For practical reasons this must be allowed (e.g. if a voter has had her credentials stolen) and in fact provides a defence against a coercer observing a ballot being cast.

\subsection{Distributed Trust}\label{ssec:distributed}
Trust in the VMV demonstrator was distributed across CES and UoS so that neither of these parties individually could be in a position to interfere undetectably with the integrity of the election.  The four mix-net nodes were distributed so that two were under the control of CES, and two under the control of UoS.  Neither party controlled a threshold of nodes, and so neither is individually able to interfere with the integrity of the mix-net.  Similar reasoning applies to the ledger: two nodes were run by CES, and two by UoS. Therefore no individual party is able to interfere with the contents of the Quorum ledger while maintaining consensus.

As discussed, the voter private keys and $\alpha$ terms are managed within VMV, outside the control of CES. This means in particular that VMV is trusted for individual verifiability. Should this trust assumption be violated, two attack scenarios would be enabled:
\begin{enumerate}
\item Given knowledge of a voter's trapdoor key, VMV or an intruder could subvert individual verifiability (by forging an $\alpha$ term opening to a different tracker); and
\item Given knowledge of a voter's signing key, VMV or an intruder could in theory modify the voter's ballot, or stuff a ballot on a non-participating voter's behalf.
\end{enumerate}

Leveraging these abilities to affect the result of an election without being detected would, however, require collusion with CES. First and foremost, the currently weak privacy guarantees mean that CES can audit VMV's behaviour and detect violations of the trust assumptions. Further, the architecture itself mitigates the attacks somewhat, relying in particular on the fact that verifiability data is immutable once committed to the ledger, unless collusion occurs between CES and VMV.

Adding a vote would be easily detectable by the non-voter whose identity was hijacked.  Modifying a vote that was honestly cast would require VMV to direct the voter to another voter's tracker showing their expected vote.  This corresponds to a clash attack, where more than one voter is directed to the same tracker so each verify the same instance of a vote, which is tallied only once rather than once per voter.  However, the $\beta$ value for the tracker is assigned to the voter before the election, and the $\alpha$ is constructed by CES and forwarded to the voter after the votes are on the bulletin board. CES is not able to generate any alternative $\alpha$ to open to a specific tracker since CES does not hold the trapdoor keys.  Hence $\alpha$ and $\beta$ for a voter are committed to the voter before VMV receives them. Further, VMV knows nothing about the identity of voters or about the votes they cast, and to attempt a clash attack would need to select a tracker blindly for the targetted voters, with a non-negligible probability that it would be detected by opening to a vote other than that cast by the voter. 

\section{Trials}\label{sec:trials}
The system was deployed in April 2019, with 4 Verificatum nodes and 4 Quorum nodes.  In each case the threshold was 3 (i.e. greater than two thirds), meaning that three nodes were required to cooperate to achieve consensus and function correctly.  CES ran two Verificatum nodes and two Quorum nodes at distinct geographic locations.  This was chosen to be below the threshold of 3 in order to ensure that CES did not possess sole control over the Verificatum or Quorum networks. However, the control of 2 nodes in each case meant that CES involvement was required to run the Verificatum and Quorum components and CES could not be excluded by a threshold of other parties.  The University of Surrey (UoS) ran one Verificatum node and one Quorum node on the Surrey campus, and ran a second one on the AWS server that was also hosting the vote verification website.

The first use of the system was with an in-house trial run of the system within Civica. This was followed by a second deployment to run the June 2019 student representatives election within the Department of Computer Science at the University of Surrey.  In both cases voters were able to verify their vote and to complete our initial questionnaire.  As a result of these two trials the system was tuned to make election setup easier, and the VMV website was simplified following voter feedback.  We also tuned the questionnaire.

The system was then used in two binding commercial ballots run by CES: a ballot in August 2019 to elect a West Midlands representative for the Royal College of Nursing, and a ballot to elect two representatives for the College of Podiatrists in October 2019. Both of these ballots were open for a period of several weeks, and in both cases the voters were invited to verify their vote after the close of the ballot, and to complete the questionnaire. The full questionnaire is given in Appendix~\ref{sec:questionnaire}.

For the initial trial, the electorate consisted of employees within a section of Civica, together with members of the project team.  The ballot asked voters to select one of a list of four chocolate bars, and the ballot was open for a week.  Following the close of the ballot, voters were offered the opportunity to verify their vote.  Those that followed the link to the verification page were then given the opportunity to complete the questionnaire.  In this ballot there were 136 votes cast, and there were 39 questionnaires started (36 completed), a proportion of 29\%.  We did not explicitly record the number of verifications performed but the requirement to verify before reaching the questionnaire gives a minimum verification rate of 29\%.

The student representatives trial consisted of two races for student representative (one for each programme), with two candidates in each race. The electorate consisted of students in a particular year on one of the degree courses that was electing a representative. Voters could only vote in the race appropriate to their degree course. In this trial there were 21 votes cast, and three questionnaires.  Although the numbers were small, this deployment was the first one external to CES and provided the confidence and feedback required to scale up to the two full trials.

The Royal College of Nursing ballot was a race between two candidates, with voters asked to select one candidate.  The electorate consisted of members of the Royal College of Nursing within the West Midlands region. There were 942 votes cast, and 503 verification checks, a verification rate of 53\%.  Of those who verified, there were 162 completed questionnaires: 17\% of those who voted.

The College of Podiatrists ballot was conducted to select two representatives from six candidates.  The electorate consisted of members of the College of Podiatrists.  Voters were asked to select two candidates from the list.  In this ballot there were 315 ballots cast, and 148 verification checks were carried out, a rate of 47\%.  There were 32 questionnaires completed: 10\% of those who voted. \\

The questionnaires were deployed "in the wild" on voters in a real binding ballot. The advantage of this approach is that we obtain responses from voters who are in exactly the situation where the system is intended to be used. However it must be borne in mind that the questionnaire was offered only to those voters who had verified their vote.  The opportunity to complete the questionnaire was therefore restricted to voters already engaged with the verification process.  This was necessary since the questionnaire is seeking feedback about the verification experience, however it is likely to bias the sample, since those that dislike technology or electronic voting are less likely to reach this stage.  The alternative would have been to carry out a ballot in a lab situation where participants are asked to carry out voting and verification tasks and then complete the questionnaire.  While this would allow for more control over obtaining a cross section of voters, the responses might not reflect how the voters would feel in a meaningful ballot.

The aim of the questionnaire was to see whether voters were happy with the VMV system and to check how they perceived, used and evaluated it. The main focus was on voters' observations that can guide future enhancement of Verifiable Voting. In particular, we were interested in voters' perceptions of VMV and factors that play important role in determining their attitude toward it. The main factors are: 

\begin{itemize}
  \item \textbf{Privacy} : "This checking system keeps my vote private."
  \item \textbf{Ease of Use} : "It was easy to check my vote."
  \item \textbf{Verifiability} : "I would check my vote next time if I could."
  \item \textbf{Appreciation Level} : "I am pleased to check my vote."
  \item \textbf{Confidence} : "Checking my vote gave me confidence that the election result is correct."
  \item \textbf{Recommendations to Others} : "I think everyone should check their vote if the facility is available."
  
\end{itemize}

The questionnaire consisted of 14 statements about the VMV system using a 6 point Likert scale, with levels of agreement ranging from Strongly Agree to Strongly Disagree; 1 open-ended question asking voters about their feedback or any further comments; and 2 categorical questions representing Gender and Age range. 

\subsection{Demographics}
\begin{figure}[h]
\includegraphics[width=0.45\textwidth]{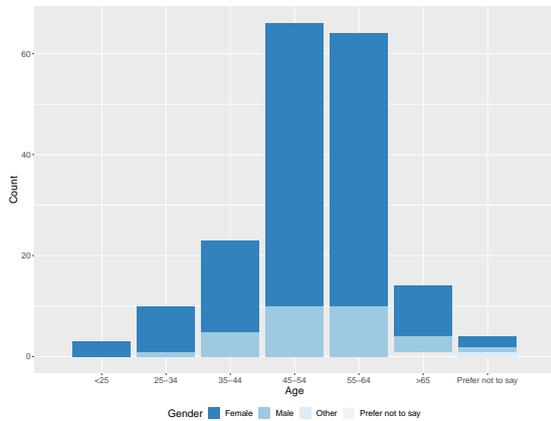}
\caption{Voters Gender and Age Distribution.} \label{fig:genderage}
\centering
\end{figure}

The data obtained after The Royal College of Nursing and The College of Podiatrists elections shows that 82.6\% of voters were females and 16.3\% were males, a gender split that is aligned with the demographics in thos two professions.  The Gender and Age distribution of voters is shown in Figure~\ref{fig:genderage}. 

\subsection{Correlations}

We used Fisher's exact test of independence to explore correlations between Gender and Age and the six constructs.  The results are presented in Table~\ref{tab:fisher} in Appendix~\ref{sec:stats}.  The main reasons for choosing Fisher's exact test are the sample size and its unbalanced distribution. The results show there is no correlation between gender and age and the six constructs, except a weak negative correlation (p=0.04) between age and appreciation level.

\begin{figure}[ht!]
\includegraphics[width=0.45\textwidth]{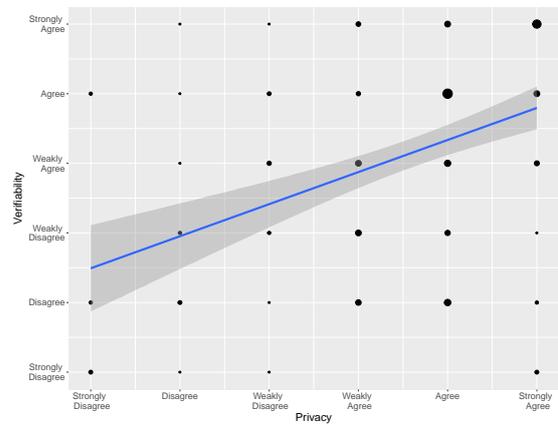}
\caption{Privacy - Verifiability Correlation.}
\label{fig:figure1}
\end{figure}

\begin{figure}[ht!]
\includegraphics[width=0.45\textwidth]{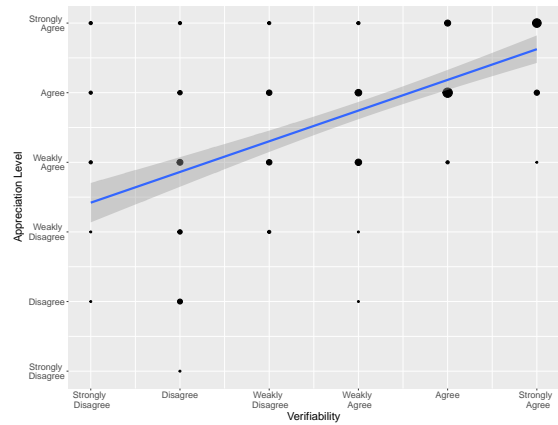}
\caption{Verifiability - Appreciation Level Correlation.}
\label{fig:figure2}
\end{figure}

\begin{figure}[ht!]
\includegraphics[width=0.45\textwidth]{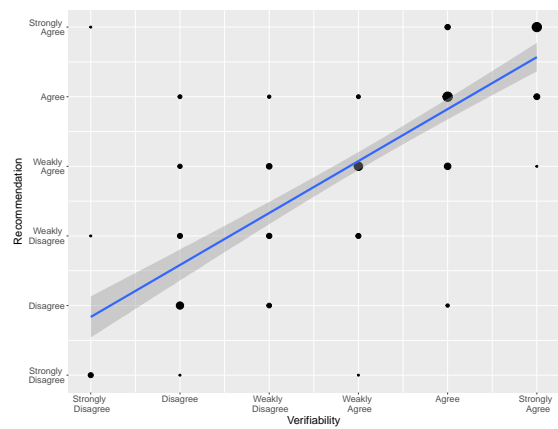}
\caption{Verifiability - Recommendation Correlation.}
\label{fig:figure3}
\end{figure}

For non-parametric correlation among construct and other variables, we use Spearman's Correlation test. The results are given in Table~\ref{tab:spearman} in Appendix~\ref{sec:stats}.  The results show the strongest correlation between Verifiability and Recommendation to others ($\rho > 0.8$), but all of Verifiability, Appreciation, Confidence, and Recommendation to others are strongly correlated ($\rho > 0.6$).  Figures~\ref{fig:figure1}, \ref{fig:figure2} and \ref{fig:figure3} illustrate some of these correlations.


\begin{figure*}[ht!]
\includegraphics[width=0.9\textwidth]{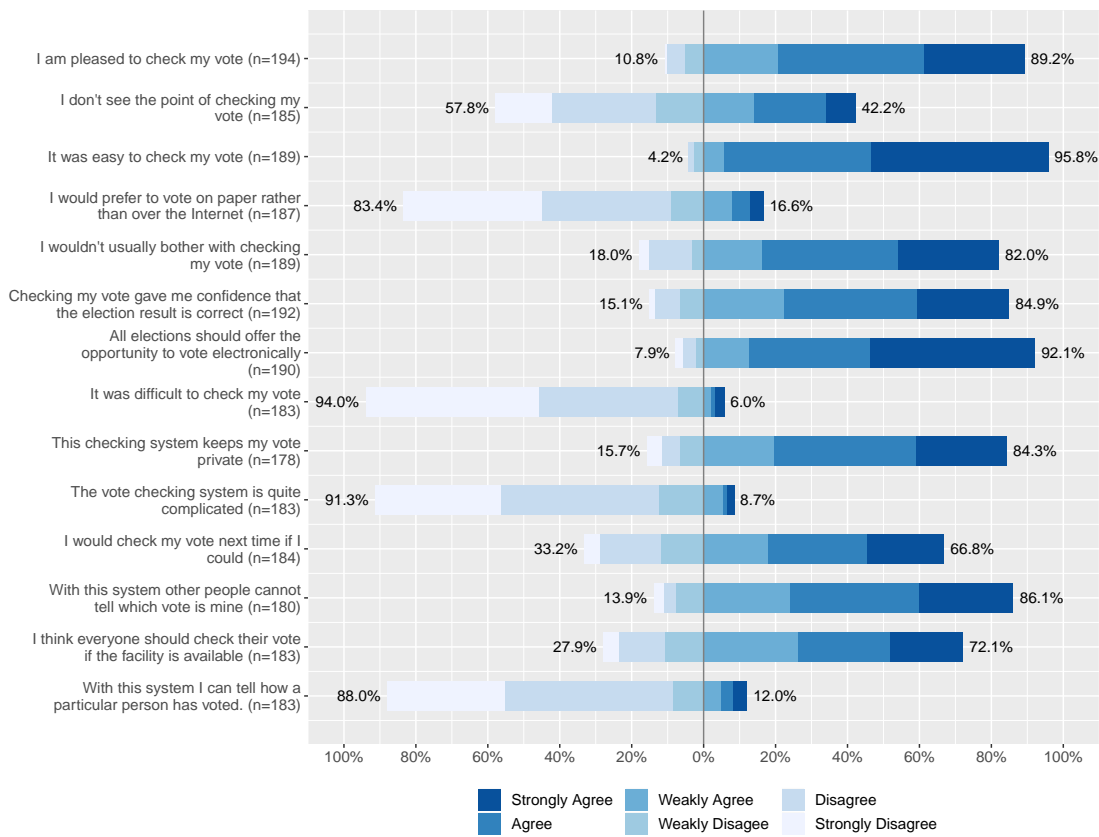}
\caption{Voters' Agreement Level Responses.}
\label{agreements}
\centering
\end{figure*}

Voters were generally happy with the opportunity to check their vote, and with the ease of doing so, and did not feel that the system was complicated. 89.2\% agreed with the statement "I am pleased to check my vote", 95.8\% agreed with the statement "It was easy to check my vote", and 94.0\% disagreed with the statement "It was difficult to check my vote". Furthermore, 84.9\% of voters expressed their agreement that the system gave them confidence that election result is correct, and 72.1\% of voters with the statement that everyone should check their vote if the facility is available. The full breakdown of answers for agreement questions is shown in Figure~\ref{agreements}.

With respect to privacy, the majority of voters recognised that the system did not reveal how a voter has voted: 84.3\% agreed that the checking system kept their vote private, 86.1\% agreed that other people could not tell which vote was theirs, and 88\% disagreed with the statement that they could tell how a particular person has voted

\begin{figure}[h]
 \centering
\includegraphics[width=0.45\textwidth]{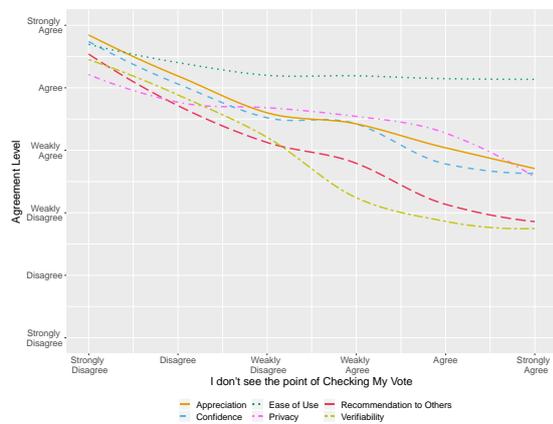}
\caption{Correlation between ``I don't see the point of checking my vote'' and other questions.} \label{fig:mannwhitney}
\centering
\end{figure}

Despite these positive findings, the question concerning the voters' comprehension of and attitude to verifiability obtained more ambiguous results that are difficult to interpret.  We had 42.2\% agreeing with the statement that "I don't see the point of checking my vote". In order to see whether this point had any influence on voters' perception of VMV, we considered the correlation between the question \textbf{"I don't see the point of checking my vote"} and \textbf{Appreciation, Ease of Use, Recommendation to others, Confidence, Privacy and Verifiability}  The correlations are shown in Figure~\ref{fig:mannwhitney}: as responders increasingly do not see the point of checking their vote, their appreciation of these other qualities all decrease.

\begin{figure*}
\includegraphics[width=0.3\textwidth]{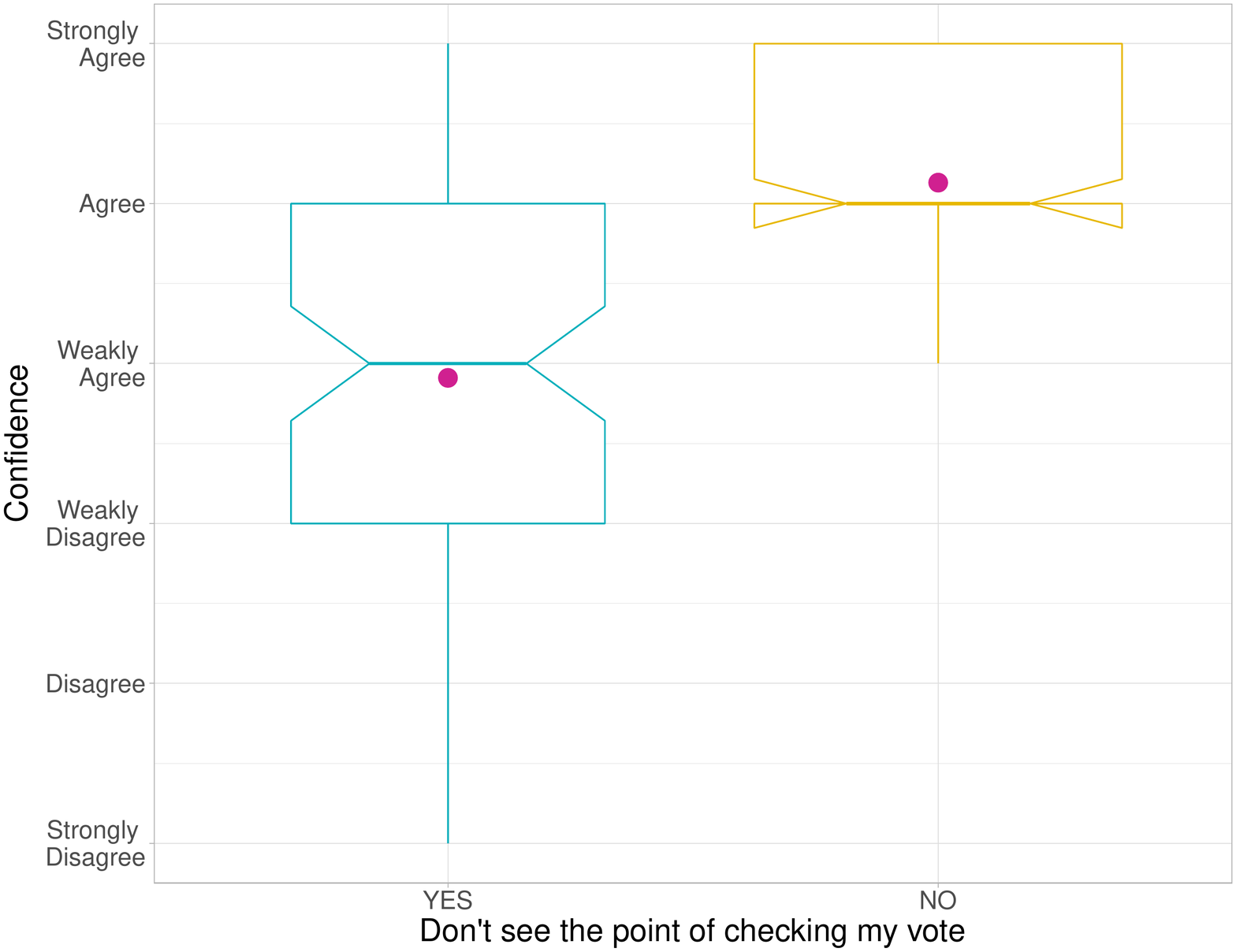}
\hspace*{0.1\textwidth}
\includegraphics[width=0.3\textwidth]{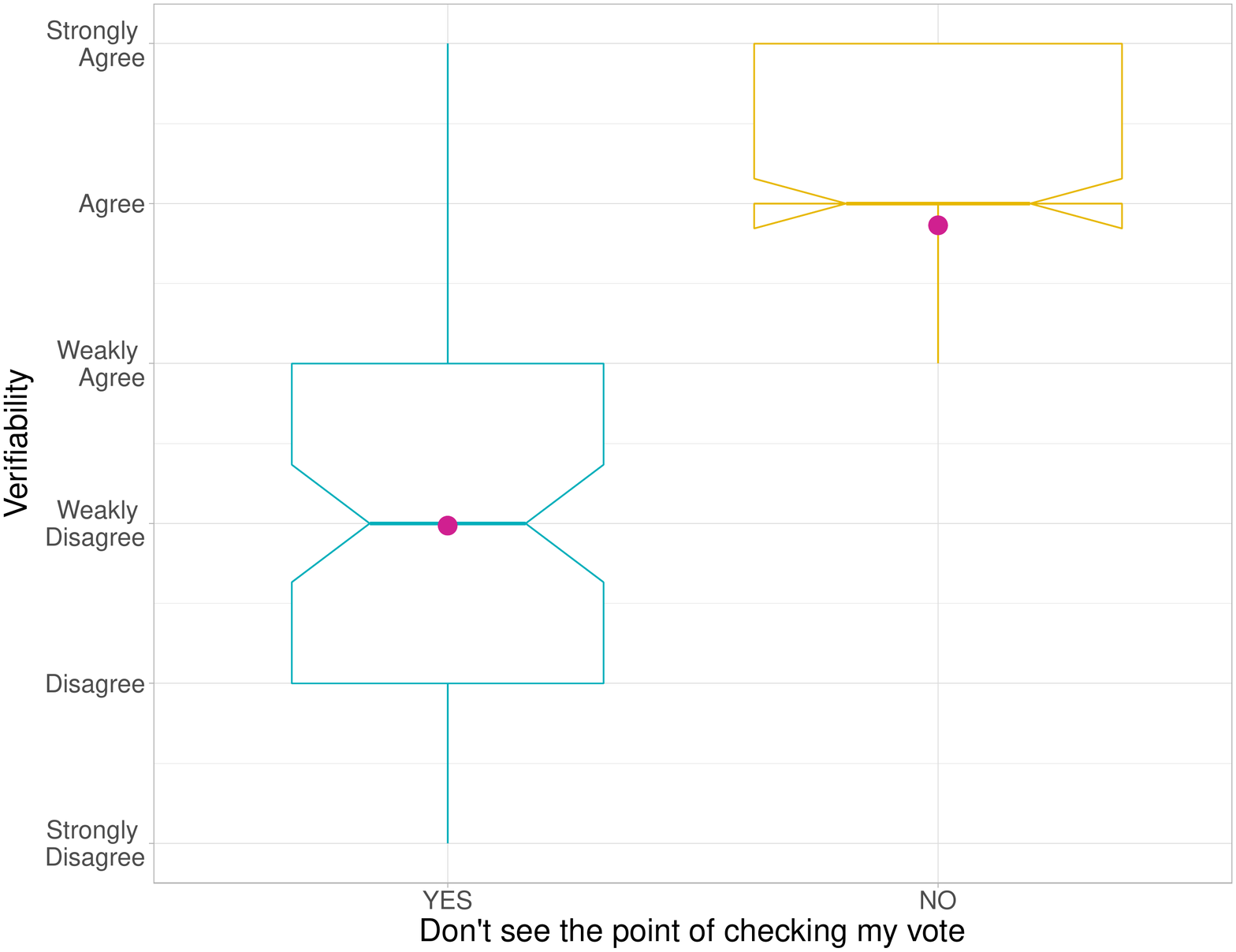}
\caption{Comparisons against Confidence and against Verifiability}
\label{fig:figure8}
\end{figure*}

\begin{figure*}[h]
\begin{center}
\includegraphics[width=0.3\textwidth]{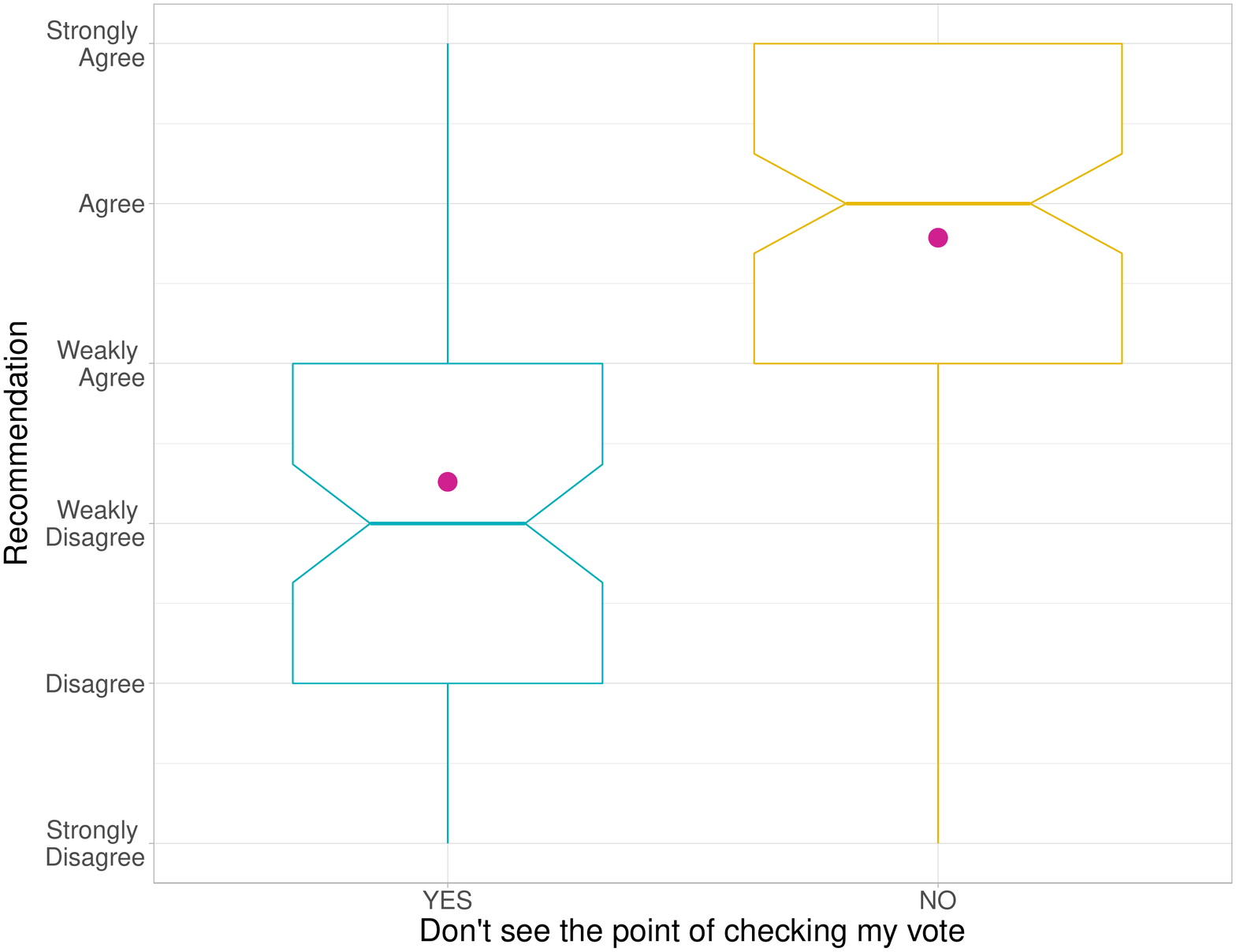}
\hspace*{0.1\textwidth}
\includegraphics[width=0.3\textwidth]{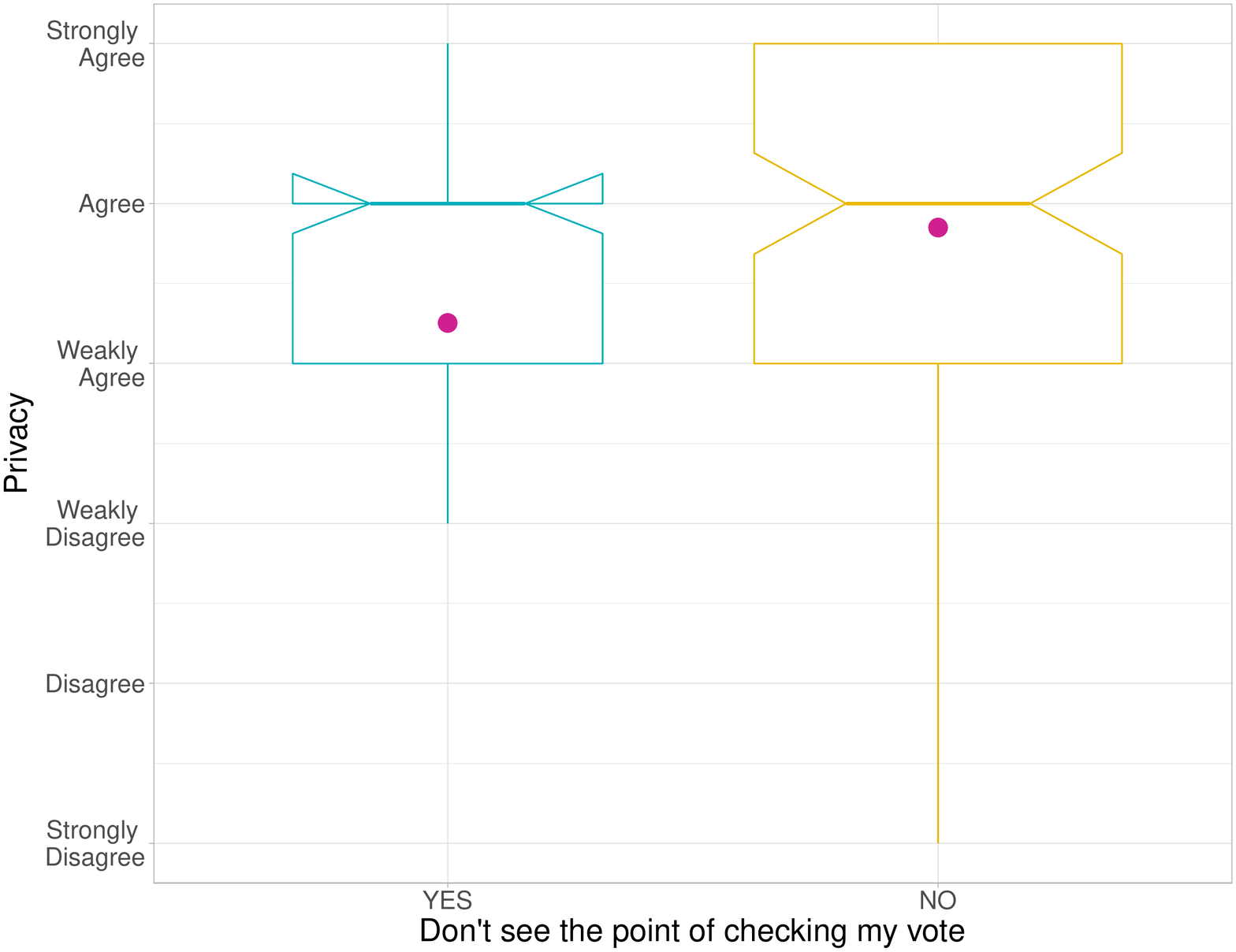}
%
%
\caption{Comparisons against Recommendation and against Privacy}
\label{fig:figure10}
\end{center}
\end{figure*}

\begin{figure*}[h]
\begin{center}
\includegraphics[width=0.3\textwidth]{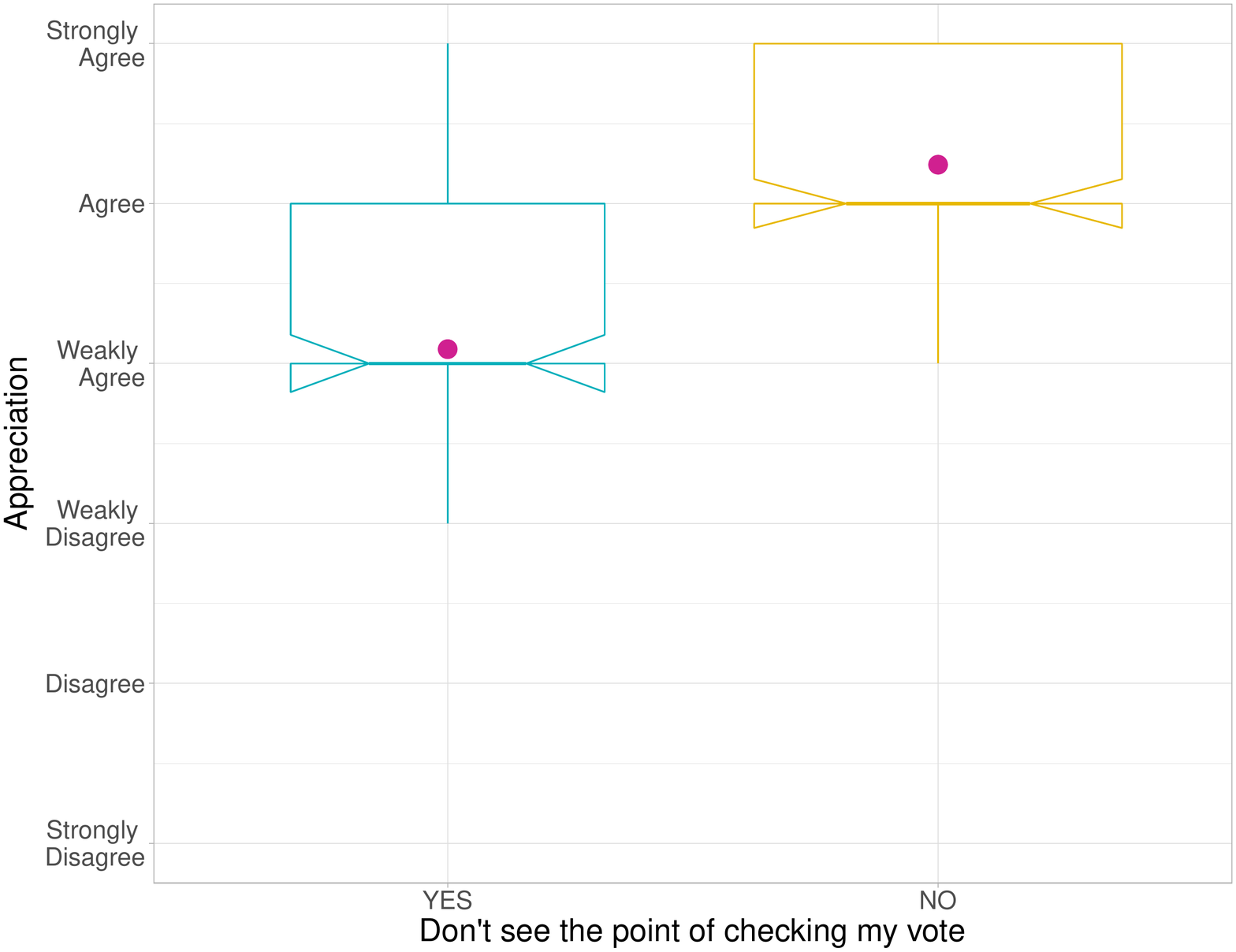}
\hspace*{0.1\textwidth}
\includegraphics[width=0.3\textwidth]{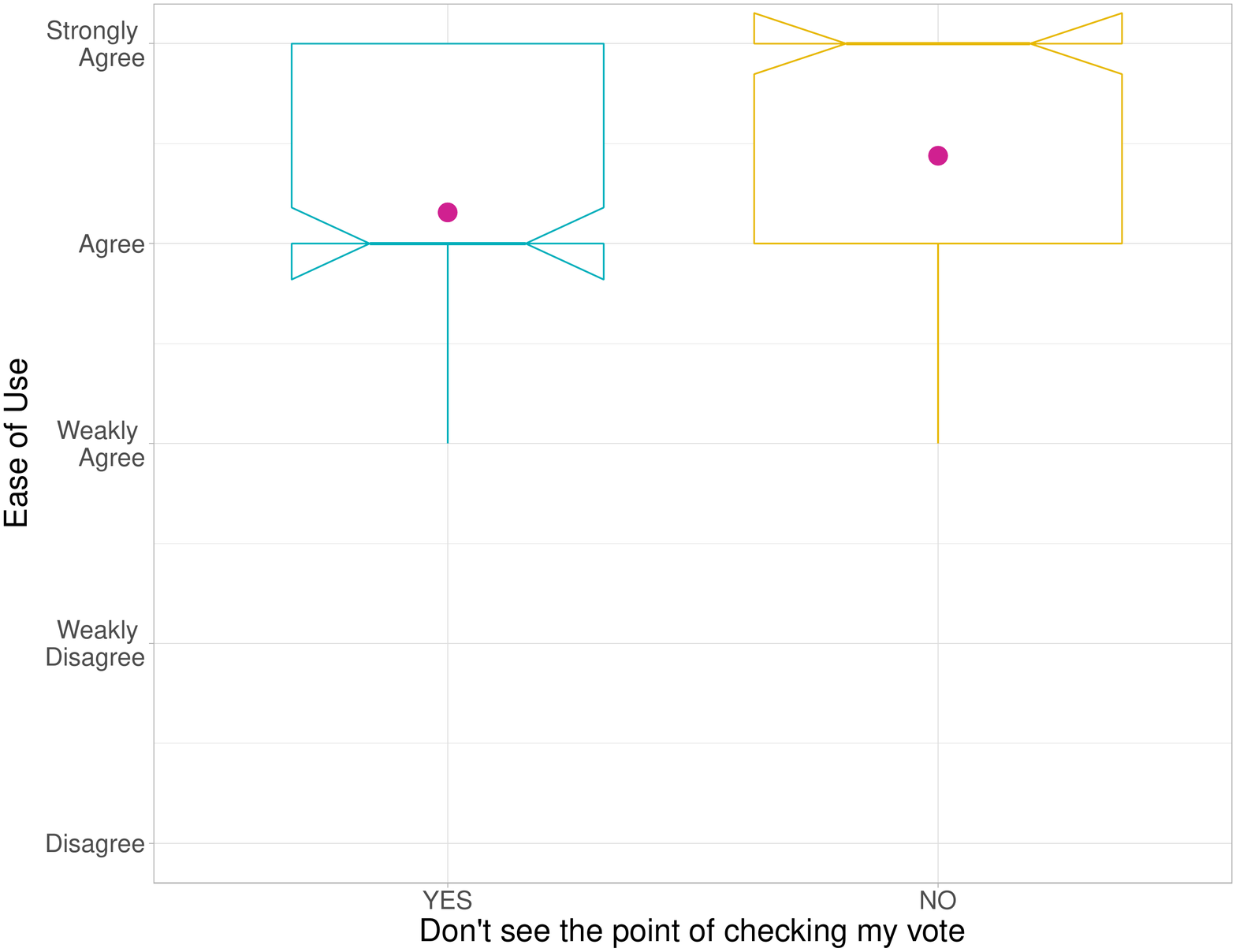}
%
%
\caption{Comparisons against Appreciation and against Ease of Use}
\label{fig:figureApprEase}
\end{center}
\end{figure*}

We evaluated each question using the Mann-Whitney test.  This test evaluates whether the voters in two groups (those that do not see the point of checking their vote, and those that do) will give similar responses to the questions, or whether they are different.  The Mann Whitney test indicated that the two groups do give different responses: in each case it rejected the null hypothesis (that the two groups are similar).  Figures~\ref{fig:figure8}, \ref{fig:figure10} and \ref{fig:figureApprEase} show the box plots for each of the questions.  They show that voters' negative feedback about the system almost always arose from those who indicated that they were unable to see the point of checking their vote. 

\subsection{Sentiment Analysis}

The questionnaire contained one open-ended question which provided an opportunity for voters to provide free form comments on any aspect of the system that they want to share. This has provided qualitative feedback giving the broad range of voters' responses to the system. The number of responses obtained by the end of the elections was 67. In order to analyze these comments "SentimentAnalysis" and "tidytext" R packages were used. \\

\begin{figure}[h]
 \centering
\includegraphics[width=0.45\textwidth]{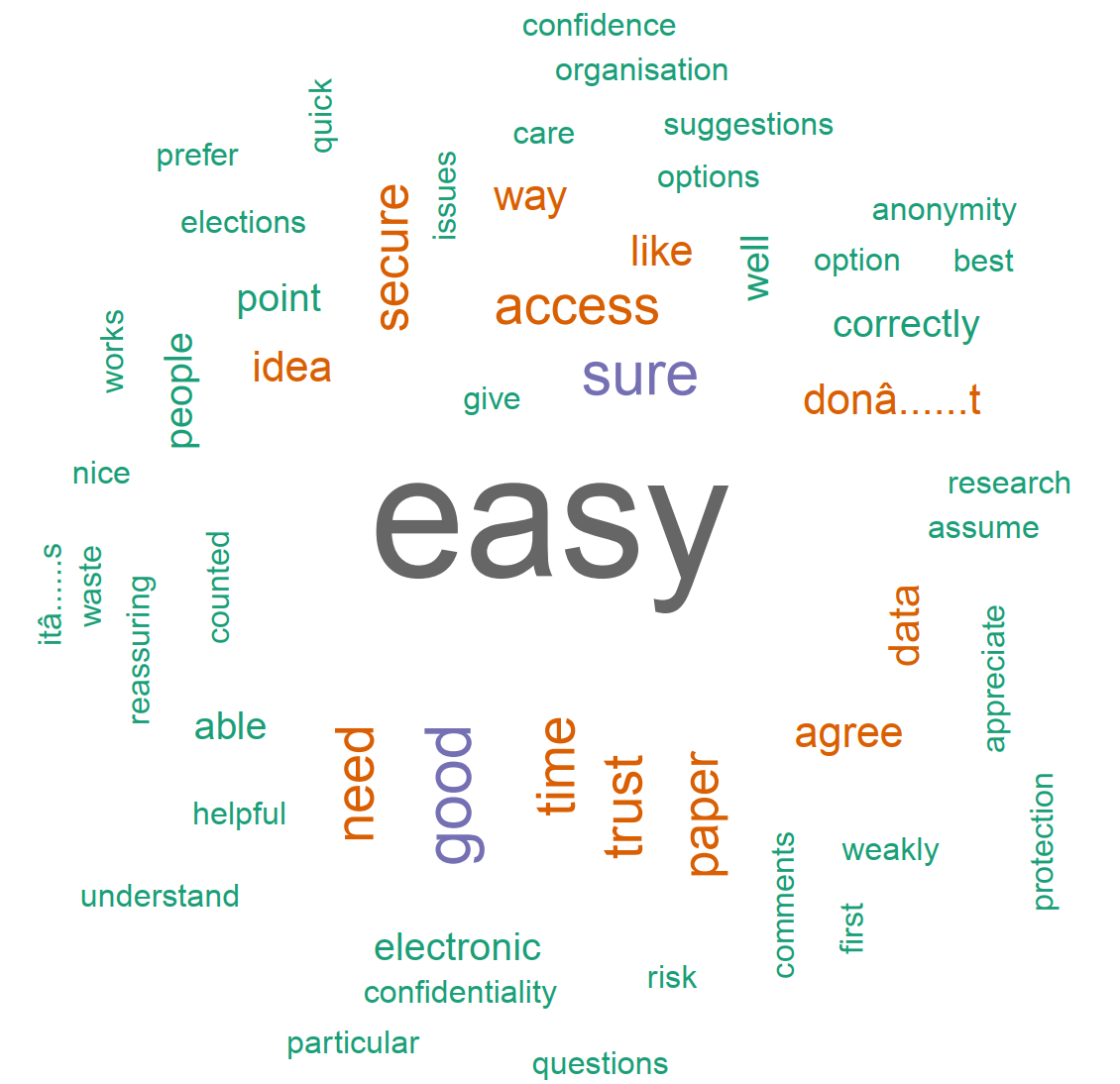}
\caption{Most Frequent Words}
\centering
\label{sentiment0}
\end{figure}

There were many positive comments around ease of use and confirming the vote, for example:
\begin{quote}
\emph{"Was easy and nice to know it was correct"} 
\end{quote}
\begin{quote}
\emph{"Very easy to use and reassuring that vote registered as I voted"}
\end{quote}
\begin{quote}
\emph{"I like and appreciate the opportunity to check my vote. In something as seemingly trustworthy as an RCN election it is a bonus. If I were voting in a General Election I would regard it as a necessity."} 
\end{quote}

Some respondents felt that the election authority should be trusted to ensure the system is secure and therefore there should be no need for verifiability, or could not see the point of verifiability:
\begin{quote}
\emph{"It isn't something I would use. I would trust that an organisation such as yours would have a secure system"}
\end{quote}
\begin{quote}
\emph{"Very easy but I don't understand the need to check. Do you feel that we don't trust the system ?"}
\end{quote}
\begin{quote}
\emph{"I don't see the point! Sorry! Once my vote is cast I assume it has been correctly processed or there are issues with the system. It tells me nothing about the robustness of the system and is a step that I would find not worthy of my time."} 
\end{quote}

Some respondents considered the question around privacy more deeply and identified some issues that would need to be answered for voters:
\begin{quote}
\emph{"Since there seems to be a permanent record of how someone votes there is risk of breach of anonymity. I would prefer no record to be held."}
\end{quote}
\begin{quote}
\emph{"As this is the first time I have used the vote checking system, the privacy of my vote is purely on a trust basis relating to data protection."}
\end{quote}
\begin{quote}
\emph{"I have no idea if this system keeps my vote private and/or that others can see how I or other people have voted."}
\end{quote}

\begin{figure*}[h]
 \centering
\includegraphics[width=0.75 \textwidth]{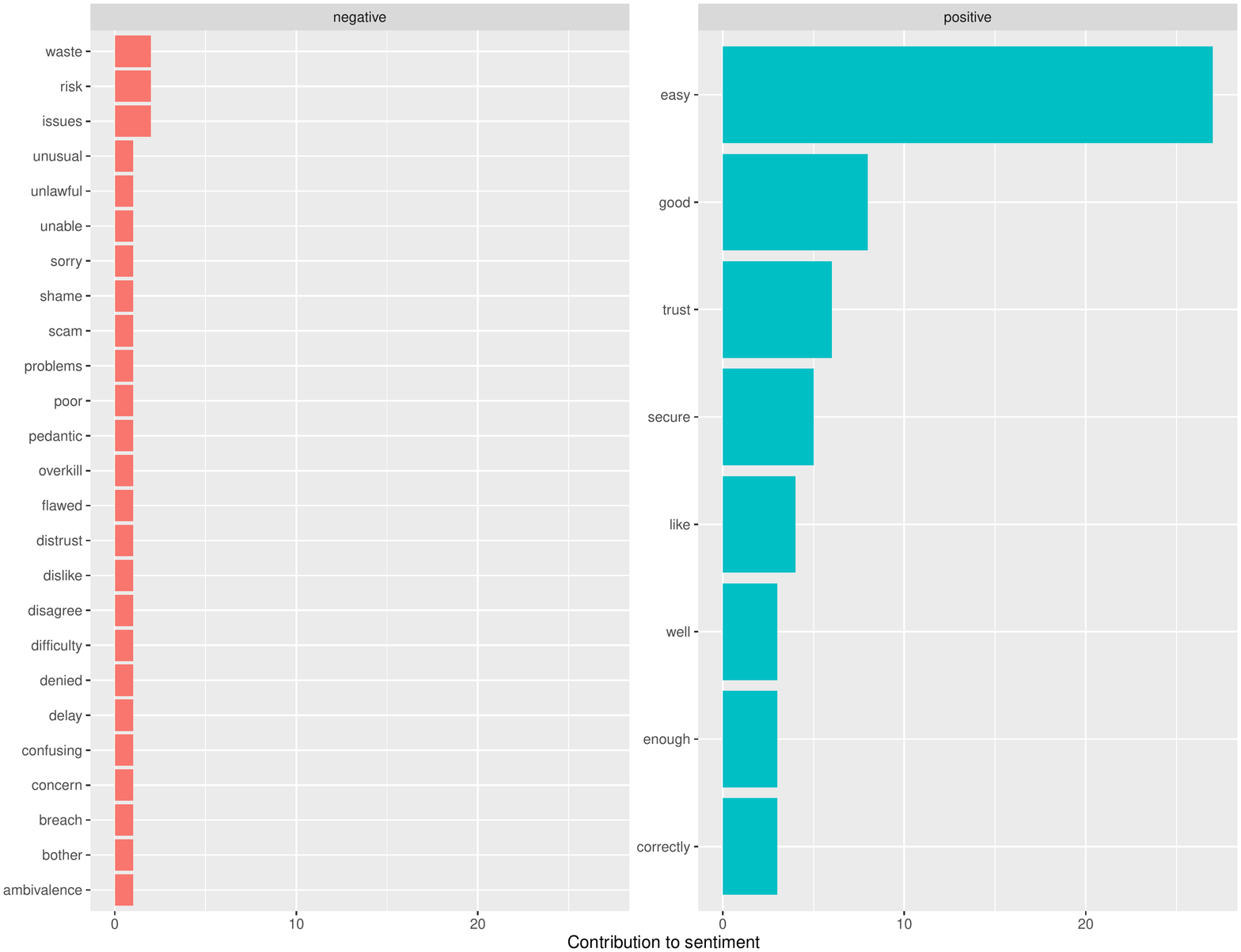}
\caption{Word Contribution to Sentiment}
\centering
\label{sentiment}
\end{figure*}

The sentiment analysis results show that 4 comments were classified as negative, 13 comments as neutral and 49 comments as positive. The word contribution to sentiment is illustrated as a word cloud in Figure~\ref{sentiment0} and also shown in Figure~\ref{sentiment}.

\section{Conclusions and Future Work} \label{sec:conclusion}

In this paper we proposed VMV (``Verify My Vote''), which adds Selene verifiability mechanisms onto a deployed internet voting system run by our commercial partner CES. Although this is an initial step it has already resulted in a system which provides stronger integrity guarantees for the CES system than it presently has, with VMV used as a simple external auditor for the conduct of specific elections.  The system provides individual and universal verifiability provided CES and VMV are not colluding to break the integrity of the election.

This initial system also has provided us with a platform for running trials ``in the wild'' on live elections to explore practical and usability issues, and to investigate open questions around voters' understanding and attitude to this approach to verifiability.  Our findings from the initial trials are that voters are able to manage the current level of verification provided to confirm that the system has correctly recorded their vote.  As further features of Selene are also introduced, these can also be investigated for usability, understanding and attitude.  Further work will need to investigate how effectively voters are able to notice mistakes in the evidence they are presented with, and also to compare with other approaches.  These questions will need to be studied in a controlled setting, where voters are aware that they are participating in a trial.  The VMV system we have developed can be used for this purpose.

As highlighted in our roadmap (Section~\ref{sec:intro:roadmap}), the VMV system is designed to be extended further towards fully trustworthy elections.
Although some of these steps are mainly engineering challenges (with further commercial constraints), there remain also research questions to explore, in particular surrounding mixed-mode voting, and coercion mitigation.
Selene can be deployed over pen-and-paper ballots~\cite{electryo}, and we show here how it can be layered on an existing voting system. It is unclear how the same level of verifiability could be offered to mixed mode voting, with some votes coming as pen-and-paper ballots (by post) and others electronically.
Further, completing Step 5 of our roadmap and adding coercion mitigation to VMV will require the development of a secure deniable channel for voters to collect the tracker openings, and a safe and usable mechanism through which non-expert voters could produce decoy trackers.

Going further towards transparency and open verifiability, there are also opportunities for more active use of the ledger.  Our current use is as a repository for information, however there is the possibility of using smart contracts to carry out some of the technical elements of verifiability automatically, for example verifying the NIZKPoKs, or in dispute resolution to carry out some of the necessary checks; this is an area of ongoing work.  There is also the opportunity to use the DLT as a public log of certificates issued by certification authorities, following the ideas of Certificate Transparency \cite{rfc6962}, to support transparency and trust in VMV.

As commented above, issues around the VMV system managing the voters' keys would obviously be more fully mitigated by allowing users to create, register and manage their own keys. This could be done progressively, allowing those voters whose desire for verifiability outweighs their trust in VMV to take control of their keys while still allowing other voters to vote smoothly while enjoying the benefits of universal verifiability, and the marginal benefits of individual verifiability checks performed by other voters.  Some of the arising challenges are pure engineering challenges. Other seem to stand as interesting research problems.

Working towards ballot privacy will require ballots to be encrypted before being sent to CES. In a first step, this could be done without signing—protecting voters' ballot privacy at a cost in universal verifiability, since CES could stuff ballots, though this would be detectable: voters who have not voted can observe that a vote has been cast against their credentials.

In practical settings, simply passing encrypted ballots to CES is a bit more problematic, as it complicates several processes, including vote cancellation and dispute resolution. In practice the system run by CES also allows for votes to be cancelled through a manual process, removing them from the database of votes. For practical reasons this must be allowed (e.g. if a voter has had her credentials stolen). The current architecture, where encrypted ballots are only collected after the election has finished, makes vote cancellations relatively easy to integrate—only the final ballot would be recorded.  We are considering ways to support vote cancellation in the context of verifiability, non-deletion of data, and how to make use of credentials in issuing replacement ballots. 
With respect to dispute resolution, supporting the current level for voters to raise challenges would require more effort.  One solution would be for CES to collect the ballot encrypted both under the election public key, and under a symmetric key controlled by the voter—with dispute resolution requiring opening. However, this increases the voter's key management burden.   

With respect to key management, even without dispute resolution mechanisms, a full deployment of Selene places
significant responsibility on the user to manage their trapdoor key (to decrypt
their trackers), and their signing key (used to sign encrypted ballot). This
burden extends beyond simply managing keys: indeed Selene requires that tracker
commitments be associated with public credentials \emph{ahead} of the election
starting, so voters would be required to create and register their credentials
ahead of time, with no opportunity to change them afterwards.

A first improvement to the current situation—--where VMV makes key management
completely opaque to voters—--would be to have VMV create and securely store
keys, but let voters extract their keys from the system to perform
cryptographic operations themselves. This is in fact quite difficult to achieve
without giving VMV a link between a voter's real identity (through an IP
address, for example) and their pseudonym, used as an index for their
credentials. The same problem arises if voters are allowed to create their own
credentials and register the corresponding public keys for use in a particular
election.

Even with such mechanisms in place, we expect that in practice most voters would not in
fact engage with the process, and instead would prefer to rely on VMV to manage
their credentials. We would envision a mechanism through which—at the end of
the registration period—VMV would generate, store and publicly register
credentials for all voters who have failed to do so themselves.

Where VMV manages the signing key, it is necessary to provide a way to obtain a signature on an encrypted ballot without any involvement from CES.  One way would be to present the voter with an encrypted ballot, and ask them to copy-paste it into the VMV website, copy-pasting the signature back. However this is unwieldy.

Alternatives that would rely on a single authentication point would be ideal,
but require solving the same challenge as that presented by key management: VMV
has the signing key, and could use it to sign the encrypted ballot, but must do
so without being able to link the pseudonym used to index the public key to any
of the voter's real-world characteristic (including their physical location).
it is worth noting here that CES could verify the signature before accepting
the encrypted and signed ballot, to prevent VMV from disenfranchising voters
through bad behaviour.

Ideally users will generate and manage their own signing keys, enabling them directly to sign the
encrypted ballots to protect against ballot stuffing. Here, it is worth noting
that CES controls the voting front-end, and must never learn the voter's
private signing key.

Mixed mode voting provides another challenge.  CES, and other established online voting solutions, must contend with the
practical constraints discussed above, but must also accept, depending on their
client, mixed mode elections, where some ballots may be cast by post or
telephone, in addition to electronic ballots. Selene has already been defined
for use with paper ballots~\cite{electryo}. However, ensuring a consistent integration
of verifiability across all voting modes with VMV as an external source of
trust is an interesting research challenge that must be solved before
deployment of private and verifiable electronic voting can become
more widespread.

The VMV demonstrator system is under development as an open-source project within the context of a research project on trustworthy voting, and has been undergoing testing and trials since April 2019.  The most natural step-change development for the system is to provide the voters with the ability to manage their own cryptographic keys, to encrypt and sign their own votes, and to obtain their own tracker, without the need to trust VMV. This would be a significant step towards a full implementation of Selene. This would most naturally be achieved through a voting app or smartcard that incorporates all of the functionality required, including the generation of key pairs and managing the interactions with the VMV system, even including the establishment of a private channel.  All of these developments bring us closer to secure and verifiable electronic voting.

\section*{Acknowledgements}

We are grateful to Peter Ryan, Peter Roenne and Marie-Laure Zollinger for discussions on the questionnaire and explanations around Selene, and to Douglas Wikstr\"{o}m for help with Verificatum. We are also grateful to Chris Fife-Schaw for advice on the questionnaire.  We are also grateful to several anonymous reviewers for their careful reading of the paper and for their suggestions for improving the paper.  This research was funded by EPSRC through the VOLT project EP/P031811/1, with support for development of VMV and integration with the CES system through the Surrey Impact Acceleration Account EP/R511791/1.


\nocite{*} 
\bibliographystyle{abbrv}           
\bibliography{bibliography}        

\begin{thebibliography}{10}

\bibitem{DBLP:conf/uss/AcemyanKBW14}
C.~Z. Acemyan, P.~T. Kortum, M.~D. Byrne, and D.~S. Wallach.
\newblock Usability of voter verifiable, end-to-end voting systems: Baseline
  data for {H}elios, {P}r{\^{e}}t {\`{a}} {V}oter, and {S}cantegrity {II}.
\newblock In {\em Electronic Voting Technology Workshop/Workshop on Trustworthy
  Elections, {EVT/WOTE} '14}. {USENIX} Association, 2014.

\bibitem{helios}
B.~Adida.
\newblock Helios: Web-based open-audit voting.
\newblock In {\em Proceedings of the 17th {USENIX} Security Symposium}, pages
  335--348, 2008.

\bibitem{aws}
{Amazon Web Services, Inc.}
\newblock Amazon web services {(AWS)} - cloud computing services.
\newblock 2019.

\bibitem{DBLP:conf/voteid/ArnaudCW13}
M.~Arnaud, V.~Cortier, and C.~Wiedling.
\newblock Analysis of an electronic boardroom voting system.
\newblock In J.~Heather, S.~A. Schneider, and V.~Teague, editors, {\em E-Voting
  and Identify - 4th International Conference, VoteID 2013}, volume 7985 of
  {\em Lecture Notes in Computer Science}, pages 109--126. Springer, 2013.

\bibitem{wombat}
J.~Ben{-}Nun, N.~Fahri, M.~Llewellyn, B.~Riva, A.~Rosen, A.~Ta{-}Shma, and
  D.~Wikstr{\"{o}}m.
\newblock A new implementation of a dual (paper and cryptographic) voting
  system.
\newblock In {\em 5th International Conference on Electronic Voting (EVOTE
  2012)}, pages 315--329, 2012.

\bibitem{benaloh}
J.~Benaloh.
\newblock Simple verifiable elections.
\newblock In {\em 2006 {USENIX/ACCURATE} Electronic Voting Technology
  Workshop}, 2006.

\bibitem{DBLP:journals/corr/BenalohRRSTV15}
J.~Benaloh, R.~L. Rivest, P.~Y.~A. Ryan, P.~B. Stark, V.~Teague, and P.~L.
  Vora.
\newblock End-to-end verifiability.
\newblock {\em \tt arXiv:1504.03778}, 2015.

\bibitem{quorum}
{Blockchain Solutions Group}.
\newblock Quorum whitepaper.
\newblock 2017.
\newblock Available from:
  https://www.blocksg.com/single-post/2017/12/27/Quorum-Whitepaper [Accessed 14
  November 2021].

\bibitem{VMV}
M.~Casey.
\newblock {VMV}: {V}erify {M}y {Vote} software.
\newblock 2020.
\newblock DOI: 10.5281/zenodo.3695909.

\bibitem{belenios}
P.~Chaidos, V.~Cortier, G.~Fuchsbauer, and D.~Galindo.
\newblock Beleniosrf: {A} non-interactive receipt-free electronic voting
  scheme.
\newblock In {\em Proceedings of the 2016 {ACM} {SIGSAC} Conference on Computer
  and Communications Security}, pages 1614--1625, 2016.

\bibitem{chaum:mixnet}
D.~Chaum.
\newblock Untraceable electronic mail, return addresses, and digital
  pseudonyms.
\newblock {\em Commun. {ACM}}, 24(2):84--88, 1981.

\bibitem{scantegrity}
D.~Chaum, R.~Carback, J.~Clark, A.~Essex, S.~Popoveniuc, R.~L. Rivest, P.~Y.~A.
  Ryan, E.~Shen, and A.~T. Sherman.
\newblock Scantegrity {II:} end-to-end verifiability for optical scan election
  systems using invisible ink confirmation codes.
\newblock In {\em {USENIX/ACCURATE} Electronic Voting Workshop, {EVT}}, 2008.

\bibitem{pretavoter}
D.~Chaum, P.~Y.~A. Ryan, and S.~A. Schneider.
\newblock A practical voter-verifiable election scheme.
\newblock In {\em Computer Security - {ESORICS} 2005, 10th European Symposium
  on Research in Computer Security}, pages 118--139, 2005.

\bibitem{Chaum:1981:UEM:358549.358563}
D.~L. Chaum.
\newblock Untraceable electronic mail, return addresses, and digital
  pseudonyms.
\newblock {\em Commun. ACM}, 24(2):84--90, Feb. 1981.

\bibitem{civitas}
M.~R. Clarkson, S.~Chong, and A.~C. Myers.
\newblock Civitas: Toward a secure voting system.
\newblock In {\em {IEEE} Symposium on Security and Privacy (S{\&}P 2008)},
  pages 354--368, 2008.

\bibitem{CSF:CulSch14}
C.~Culnane and S.~A. Schneider.
\newblock A peered bulletin board for robust use in verifiable voting systems.
\newblock In {\em {IEEE} 27th Computer Security Foundations Symposium, {CSF}},
  pages 169--183, 2014.

\bibitem{ers}
{Electoral Reform Services Ltd}.
\newblock {ERS} - leading provider for elections | ballots | online voting |
  {AGMs}.
\newblock 2019.

\bibitem{elgamal}
T.~{Elgamal}.
\newblock A public key cryptosystem and a signature scheme based on discrete
  logarithms.
\newblock {\em IEEE Transactions on Information Theory}, 31(4):469--472, July
  1985.

\bibitem{ethereum}
Ethereum.
\newblock Home | ethereum.
\newblock 2019.
\newblock Available from: https://ethereum.org/, [Accessed 14 Nov 2021].

\bibitem{solidity}
Ethereum.
\newblock Solidity -- solidity 0.4.21 documentation.
\newblock 2019.
\newblock https://github.com/ethereum/solidity [Accessed 21 November 2019].

\bibitem{fips:186:4}
{FIPS 186-4}.
\newblock Digital signature standard {(DSS)}.
\newblock Standard, National Institute of Standards and Technology, July 2013.

\bibitem{DBLP:conf/voteid/Gjosteen11}
K.~Gj{\o}steen.
\newblock The {N}orwegian internet voting protocol.
\newblock In A.~Kiayias and H.~Lipmaa, editors, {\em E-Voting and Identity -
  Third International Conference, VoteID 2011}, volume 7187 of {\em Lecture
  Notes in Computer Science}, pages 1--18. Springer, 2011.

\bibitem{DBLP:journals/adt/GjosteenL16}
K.~Gj{\o}steen and A.~S. Lund.
\newblock An experiment on the security of the {N}orwegian electronic voting
  protocol.
\newblock {\em Ann. des T{\'{e}}l{\'{e}}communications}, 71(7-8):299--307,
  2016.

\bibitem{go}
Google.
\newblock The go programming language.
\newblock 2019.
\newblock Available from: https://golang.org/ [Accessed 14 November 2021].

\bibitem{quorum-web}
GoQuorum.
\newblock Home | quorum.
\newblock 2021.
\newblock Available from: https://www.goquorum.com/ [Accessed 14 November
  2021].

\bibitem{rubyonrails}
D.~H. Hansson.
\newblock Ruby on rails | a web-application framework that includes everything
  needed to create database-backed web applications according to the
  model-view-controller {(MVC)} pattern.
\newblock 2019.
\newblock Available from: https://rubyonrails.org/ [Accessed 14 November 2021].

\bibitem{DBLP:conf/voteid/HeibergMVW16}
S.~Heiberg, T.~Martens, P.~Vinkel, and J.~Willemson.
\newblock Improving the verifiability of the {E}stonian internet voting scheme.
\newblock In R.~Krimmer, M.~Volkamer, J.~Barrat, J.~Benaloh, N.~J. Goodman,
  P.~Y.~A. Ryan, and V.~Teague, editors, {\em Electronic Voting - First
  International Joint Conference, E-Vote-ID}, volume 10141 of {\em Lecture
  Notes in Computer Science}, pages 92--107. Springer, 2016.

\bibitem{DBLP:conf/ev/HeibergW14}
S.~Heiberg and J.~Willemson.
\newblock Verifiable internet voting in {E}stonia.
\newblock In R.~Krimmer and M.~Volkamer, editors, {\em 6th International
  Conference on Electronic Voting: Verifying the Vote, {EVOTE}}, pages 1--8.
  {IEEE}, 2014.

\bibitem{DBLP:conf/voteid/HelbachS07}
J.~Helbach and J.~Schwenk.
\newblock Secure internet voting with code sheets.
\newblock In {\em E-Voting and Identity, First International Conference,
  {VOTE-ID}}, pages 166--177, 2007.

\bibitem{gdpr}
{Information Commissioner's Office}.
\newblock Guide to the general data protection regulation - {GOV.UK}.
\newblock 2018.
\newblock Available from:
  https://www.gov.uk/government/publications/guide-to-the-general-data-protection-regulation
  [Accessed 14 November 2021].

\bibitem{DBLP:conf/fc/IovinoRRR17}
V.~Iovino, A.~Rial, P.~B. R{\o}nne, and P.~Y.~A. Ryan.
\newblock Using {S}elene to verify your vote in {JCJ}.
\newblock In M.~Brenner, K.~Rohloff, J.~Bonneau, A.~Miller, P.~Y.~A. Ryan,
  V.~Teague, A.~Bracciali, M.~Sala, F.~Pintore, and M.~Jakobsson, editors, {\em
  Financial Cryptography and Data Security - {FC} 2017 International Workshops,
  WAHC, BITCOIN, VOTING, WTSC, and TA}, volume 10323 of {\em Lecture Notes in
  Computer Science}, pages 385--403. Springer, 2017.

\bibitem{DBLP:journals/compsec/JoaquimFR13}
R.~Joaquim, P.~Ferreira, and C.~Ribeiro.
\newblock {EVIV:} an end-to-end verifiable internet voting system.
\newblock {\em Computers {\&} Security}, 32:170--191, 2013.

\bibitem{DBLP:conf/voteid/KhazaeiW17}
S.~Khazaei and D.~Wikstr{\"{o}}m.
\newblock Return code schemes for electronic voting systems.
\newblock In R.~Krimmer, M.~Volkamer, N.~B. Binder, N.~Kersting, O.~Pereira,
  and C.~Sch{\"{u}}rmann, editors, {\em Electronic Voting - Second
  International Joint Conference, E-Vote-ID}, volume 10615 of {\em Lecture
  Notes in Computer Science}, pages 198--209. Springer, 2017.

\bibitem{DBLP:conf/interact/KulykHRV19}
O.~Kulyk, J.~Henzel, K.~Renaud, and M.~Volkamer.
\newblock Comparing "challenge-based" and "code-based" internet voting
  verification implementations.
\newblock In D.~Lamas, F.~Loizides, L.~E. Nacke, H.~Petrie, M.~Winckler, and
  P.~Zaphiris, editors, {\em Human-Computer Interaction - {INTERACT} 2019 -
  17th {IFIP} {TC} 13 International Conference}, volume 11746 of {\em Lecture
  Notes in Computer Science}, pages 519--538. Springer, 2019.

\bibitem{DBLP:conf/csfw/KustersMST16}
R.~K{\"{u}}sters, J.~M{\"{u}}ller, E.~Scapin, and T.~Truderung.
\newblock s{E}lect: A lightweight verifiable remote voting system.
\newblock In {\em {IEEE} 29th Computer Security Foundations Symposium, {CSF}},
  pages 341--354. {IEEE} Computer Society, 2016.

\bibitem{rfc6962}
B.~Laurie, A.~Langley, and E.~K{\"{a}}sper.
\newblock Certificate transparency.
\newblock {\em {RFC}}, 6962:1--27, 2013.

\bibitem{bouncycastle}
{Legion of the Bouncy Castle Inc.}
\newblock The legion of the bouncy castle java cryptography {APIs}.
\newblock 2019.
\newblock Available from: https://www.bouncycastle.org/ [Accessed 14 November
  2021].

\bibitem{java}
Oracle.
\newblock Java | oracle.
\newblock 2019.
\newblock Available from: https://www.java.com/ [Accessed 21 November 2021].

\bibitem{Ramchen:2010:PSA:1924892.1924904}
K.~Ramchen and V.~Teague.
\newblock Parallel shuffling and its application to {P}r\^{e}t \`{a} {V}oter.
\newblock In {\em Proceedings of the 2010 International Conference on
  Electronic Voting Technology/Workshop on Trustworthy Elections}, EVT/WOTE'10,
  pages 1--8. USENIX Association, 2010.

\bibitem{electryo}
P.~Roenne, P.~Y. Ryan, and M.-L. Zollinger.
\newblock Electryo, in-person voting with transparent voter verifiability and
  eligibility verifiability.
\newblock In {\em Third International Joint Conference on Electronic Voting
  E-Vote-ID 2018: TUT Press Proceedings}, 2018.

\bibitem{selene}
P.~Y.~A. Ryan, P.~B. R{\o}nne, and V.~Iovino.
\newblock Selene: Voting with transparent verifiability and
  coercion-mitigation.
\newblock In {\em Financial Cryptography and Data Security - {FC} 2016
  International Workshops, BITCOIN, VOTING, and WAHC}, pages 176--192, 2016.

\bibitem{pgd}
P.~Y.~A. Ryan and V.~Teague.
\newblock Pretty good democracy.
\newblock In {\em Security Protocols XVII, 17th International Workshop,
  Cambridge, UK, April 1-3, 2009. Revised Selected Papers}, pages 111--130,
  2009.

\bibitem{sakokilian:mixnet}
K.~Sako and J.~Kilian.
\newblock Receipt-free mix-type voting scheme.
\newblock In L.~C. Guillou and J.-J. Quisquater, editors, {\em Advances in
  Cryptology --- EUROCRYPT '95}, 1995.

\bibitem{swisspost}
{Swiss Post}.
\newblock Protocol of the {S}wiss {P}ost voting system.
\newblock Online.
\newblock Retrieved 19/05/2021.

\bibitem{political:constitutional:2015}
{The Political and Constitutional Reform Committee}.
\newblock Voter engagement in the {UK}: Follow up - political and
  constitutional reform.
\newblock 2015.
\newblock Available from:
  https://publications.parliament.uk/pa/cm201415/cmselect/
  cmpolcon/938/93802.htm [Accessed 14 November 2021].

\bibitem{verificatum}
D.~Wikstr{\"{o}}m.
\newblock Verificatum.
\newblock 2021.
\newblock Available from: https://www.verificatum.com/ [Accessed 14 November
  2021].

\end{thebibliography}

%

\appendix

\section{Appendix}

\subsection{Statistical Analyses} \label{sec:stats}

\subsubsection{Fisher's test}

We used Fisher's exact test of independence to explore correlations between Gender and Age and the six constructs. Fisher's exact test was chosen as suited to the sample size and its unbalanced distribution. 


\begin{table*}[h]
\caption{Fisher's exact test} \label{tab:fisher}
\begin{tabular}{lcccccc}
\hline
& \multicolumn{6}{c}{Construct Variables}                                                        \\ \cline{2-7} 
                  & \textit{Privacy} & \textit{Ease of Use} & \textit{Verifiability} & \textit{Appreciation Level} & \textit{Confidence} & \textit{Rec. to Others} \\ \hline
Gender        & 0.802        & 0.849            & 0.176          & 0.090 
& 0.471 & 0.477     \\ \hline
Age             & 0.845       & 0.701             & 0.643          & 0.036 & 0.670 & 0.872     \\ \hline
%
%
\end{tabular}
\end{table*}

The results show that there is no correlation between Gender or Age and the six constructs except a weak correlation between Age and Appreciation Level (p = 0.036). 

\subsubsection{Spearman's Correlation test}

For non-parametric correlation among construct and other variables, we use Spearman's Correlation test. The strongest correlation was between Verifiability and Recommendation to Others ($\rho > 0.8$), but all of Verifiability, Appreciation, Confidence, and Recommendation to Others are strongly correlated (Spearman's $\rho > 0.6$ in all cases), and all with $p$ values less than $0.01$ except Ease of Use vs Privacy ($p = 0.03$), giving high confidence in these correlations. 

\begin{table*}[t]
\begin{tabular}{rcccccc}
\hline
& \multicolumn{6}{c}{Independent Variables}                                                           \\ \cline{2-7} 
                  & Privacy & Ease of Use & Verifiability & Appreciation & Confidence & Recommendation \\ \cline{2-7} 
Privacy          & --- & & & & & \\ \hline 
Ease of Use    &  0.17 & --- & & & & \\ \hline
Verifiability     & 0.40 & 0.23 & --- & & & \\ \hline
Appreciation     & 0.41 & 0.26 & 0.65 & --- & & \\ \hline
Confidence     & 0.66 & 0.37 & 0.69 & 0.66 & --- & \\ \hline
Recommendation     & 0.45 & 0.45 & 0.80 & 0.61 & 0.68 & --- \\ \hline
\end{tabular}
\caption{Spearman's Rank Correlation $\rho$ values} \label{tab:spearman}
\end{table*}

\subsection{Vote Checking Questionnnaire} \label{sec:questionnaire}

The questionnaire used in this study is given below. Participants were asked to state the extent to which they agreed or disagreed with a selection of statements, on a six point scale.  This was followed by an opportunity for any further comments in free-form.  Finally, demographic information was requested.

The questionnaire was made available to participants on the web page where they checked their votes, with the intention of evaluating their experience and understanding following the verification process they had just been through.  It was not offered to voters who chose not to verify their vote.

The questionnaire was presented as a succession of screens as follows:


\todo[color=white,inline,caption={}]{You are being invited to participate in a research study titled  "Evaluation of Vote Checking with the VMV System". This study is being led by Professor Steve Schneider at the University of Surrey.    

The purpose of this research study is to understand your experience of the vote checking system, and your opinions about electronic voting more generally. This questionnaire will take approximately 5 minutes to complete. 

Your participation in this study is entirely voluntary and you are free to withdraw at any time. You are free to omit any question. Your answers in this study will remain confidential.} 
 

{\em [For each of the next two screens a 6-point scale was provided for each of the statements: Strongly agree;	Agree;	Weakly agree;	Weakly disagree;	Disagree;	Strongly disagree.] }

\todo[color=white,inline,caption={}]{%
The following questionnaire relates to the vote checking system you have just experienced.

You will be provided with a number of statements. For each statement you should indicate to what extent you agree or disagree with the statement. 

\begin{itemize}
\item 
I was pleased with the opportunity to check my vote.								
\item 
I don’t see the point of checking my vote.								
\item 
It was easy to check my vote.								
\item 
If given the choice I would prefer to vote on paper rather than over the Internet.								
\item 
I wouldn’t usually bother with checking my vote.								
\item 
Checking my vote gave me confidence that the election result is correct.								
\item 
All elections should offer the opportunity to vote electronically.
\end{itemize}
}

\todo[color=white, inline, caption={}]{
\begin{itemize}
\item It was difficult to check my vote.								
\item This checking system keeps my vote private.								
\item The vote checking system is quite complicated.								
\item I would check my vote next time if I could.								
\item With this system other people cannot tell which vote is mine.								
\item I think everyone should check their vote if the facility is available.								
\item With this system I can tell how a particular person has voted.								
\end{itemize}
}


\todo[color=white,inline]{Please provide any further comments you may have.  For example, what do you think of the vote checking system?  Do you have any suggestions for how it could be improved?  How easy was it to use?  Any other comments?}



\todo[color=white,inline, caption={}]{These Questions are about you:

What is your gender?

\begin{itemize}
\item
Female
\item
Male
\item
Other
\item
Prefer not to say
\end{itemize}

What is your age range?

\begin{itemize}
\item
Under 25
\item
25-34
\item
35-44
\item
45-54
\item
55-64
\item
65 or over
\item
Prefer not to say
\end{itemize}

Would you be willing to participate in a follow-up interview conducted by University of Surrey as part of the research study? This is entirely optional. If you are willing to take part then please provide your email address. This will only be used to contact you in connection with this study, it will not be used for any other purpose.
}


\todo[color=white,inline]{You have the right to withdraw from this study, but by submitting the questionnaire you are agreeing to participate.

This research study is led by Professor Steve Schneider.  If you have any further comments or enquiries then please contact him at s.schneider@surrey.ac.uk. 
}

\todo[color=white,inline]{
We thank you for your time spent taking this survey.  
Your response has been recorded.}

\end{document}